\documentclass[twocolumn,amsmath,amssymb,superscriptaddress]{revtex4}
\usepackage{dcolumn}
\usepackage{bm}
\usepackage{amsmath}
\usepackage{amsfonts}
\usepackage{graphics}
\usepackage{graphicx}

\begin{document}

%
\title{Methods for detrending success metrics to account for

inflationary and deflationary factors}
\author{Alexander M. Petersen\footnotemark[1] }
\affiliation{Center for Polymer Studies and Department of Physics, Boston University, Boston, Massachusetts 02215, USA}
\author{Orion Penner}
\affiliation{Complexity Science Group, Department of Physics and Astronomy, University of Calgary, Calgary, Alberta T2N
1N4, Canada}
\author{H. Eugene Stanley}
\affiliation{Center for Polymer Studies and Department of Physics, Boston University, Boston, Massachusetts 02215, USA}
\date{\today}

\begin{abstract} There is a long standing debate over how to objectively compare the career achievements of
professional athletes from different historical eras.  Developing an objective approach will be of particular importance
over the next decade as {\it Major League Baseball} (MLB) players from the ``steroids era''  become eligible for Hall of Fame
induction.  Some experts are calling for {\it asterisks (*)} to be placed next to the career statistics of athletes
found guilty of using performance enhancing drugs (PED).  Here we address this issue, as well as the general
problem of comparing statistics from distinct eras, by detrending the seasonal statistics of professional baseball
players.
We detrend player statistics by normalizing achievements to seasonal averages, which accounts for changes in
relative player ability resulting from both
exogenous and endogenous factors, such as talent dilution from expansion, equipment and training improvements, as well
as PED.  In this paper we compare the probability density function (pdf) of detrended career statistics to the pdf of
raw career statistics for five  statistical categories --- hits (H), home runs (HR), runs batted in (RBI), wins (W) and
strikeouts (K) ---  over the 90-year period 1920-2009.  We find that the functional form of these pdfs are stationary under detrending.
This stationarity implies that the statistical regularity observed in the right-skewed distributions for longevity and
success in professional baseball arises from both the wide range of intrinsic talent among athletes and the underlying 
nature of competition.
 Using this simple detrending technique, we examine the top 50 all-time careers for H, HR, RBI, W and K.  While the
traditional order tends to be maintained for H and W, this method reveals some interesting careers that emerge from the
traditional ranks, especially in the cases of HR, RBI and K.  We fit the  pdfs for career success by the Gamma distribution in order to
calculate objective benchmarks based on extreme statistics which can be used for the identification of extraordinary careers.
\end{abstract}

\maketitle

\section{Introduction}
 Quantitative measures for  success are important for comparing both individual and group accomplishments \cite{Plos0}, often
achieved in different time periods. 
However, the evolutionary nature of competition results in a non-stationary rate of success, that makes comparing
accomplishments across 
time statistically biased. The analysis of sports records reveals  that the interplay between technology and
ecophysiological limits results  in a complex rate of record progression \cite{Plos1,Plos2,Plos3}. Since record events
correspond to  extreme achievements, 
a natural follow-up question is: How does the success rate of more common achievements evolve in competitive arenas?
To answer this question, we analyze the evolution of success, and the resulting implications on metrics for career
success, for all   Major League Baseball (MLB) players over
the entire history of the game. We use concepts from statistical physics to identify statistical regularity in
success, ranging from common to  extraordinary careers.

\footnotetext[1]{ Corresponding Author: Alexander M. Petersen \newline
\emph{E-mail}: amp17@physics.bu.edu}

\subsection{Baseball} The game of baseball has a rich history, full of scandal, drama and controversy \cite{KB99}. 
Indeed, the importance of baseball in American culture is evident in the game's longevity, having survived the Great
Depression, two World Wars, racial integration, free agency, and multiple player strikes.  When comparing players from
different time periods it is often necessary to rely purely on statistics, due to the simple fact that Major League
Baseball's  130+ year history spans so many human generations, extending back to a time period  before television
and even before public radio. 

Luckily, due to the invention of the box score very early in the evolution of the game, baseball has an extremely rich
statistical history. When comparing two players, objectively determining who is better should be as straightforward
as comparing their
statistics.  However, the results of such a naive approach can be unsatisfying.  This is due to the fact that the
history of professional baseball is typically thought of as a collection of ill-defined, often overlapping eras, such as
the ``deadball" era,  the ``liveball" era, and recently, the ``steroids" era of the 1990's and 2000's.  As a result,
many careers span at least two such eras.  


The use of statistics, while invaluable to any discussion or argument, requires proper contextual interpretation.  This
is especially relevant when dealing with the comparison of baseball careers from significantly different periods. 
Among common fans, there will always be  arguments and intergenerational debates.  Closely related  to these  debates,
but on a grander stage, is the election process for elite baseball players into the great bastion of baseball history,
the National Baseball Hall of Fame (HOF).  In particular, an unbiased method for quantifying career achievement 
would be extremely useful in addressing  two issues which are on the horizon for the HOF: 

(i) How  should the   HOF reform the election procedures of the veterans committee, which is a special committee 
responsible for the retroactive induction of
players who were initially overlooked during their tenure on the HOF
ballot.  Retroactive induction is the only way a player can be inducted into the HOF once their voting tally drops below
a 5\% threshold, after which they are not considered on future ballots.  Closely related to induction through the
veterans committee is the induction of deserving African American players who were not allowed
  to compete in MLB prior to 1947, but who excelled in the Negro Leagues, a separate baseball league established for  ``players of color.''  In 2006, the HOF welcomed seventeen Negro
Leaguers in a special induction to the HOF.

(ii) How should the HOF deal with players from the ``steroid" era (1990's - 2000's) when they become eligible for HOF
induction. The  {\it Mitchell Report} \cite{mitchell} revealed that more than 5\% of players in 2003 were using PED.
Hence,  is right to celebrate the accomplishments of players guilty of using PED more than the accomplishments of
the players who were almost as good and were  not guilty of using PED?  
Similarly,  how can we fairly assess player accomplishments from the steroids era without discounting the
accomplishments of innocent players?

Here we address the era dependence of player statistics in a straightforward way.  We develop a 
quantitative method to ``detrend" seasonal statistics by the corresponding league-wide average. As a result, we 
normalize  accomplishments across {\it all} possible performance factors inherent to a given time period. Our results
provide an unbiased and statistically robust appraisal of career achievement, which can  be extended to other sports and
other professions where metrics for success are available.

\begin{figure}
\centering{\includegraphics[width=0.45\textwidth]{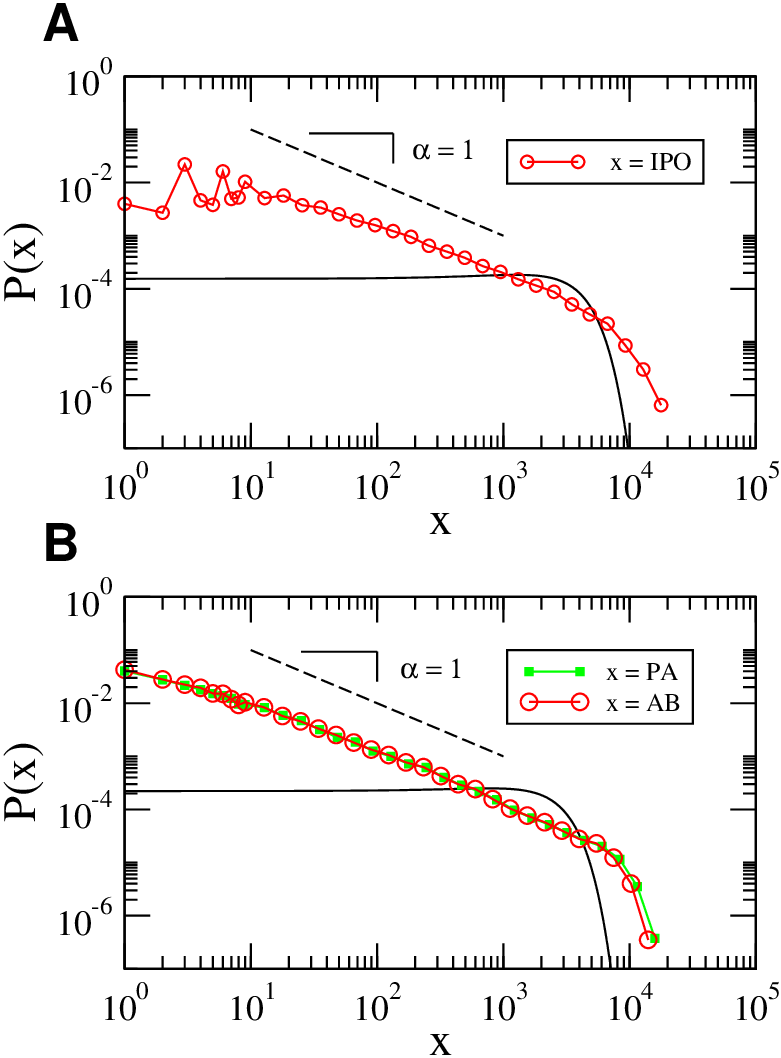}}
 \caption{   \label{long}  The  probability density functions (pdfs) of player longevity demonstrate the wide range of careers. 
We define longevity as {\bf (A)} the number of outs-pitched (IPO) for
pitchers and {\bf (B)} the
 number of at-bats (AB)
for batters. {\bf (B)} For comparison, we also plot the pdf for plate appearances (PA) for batters, which is not
significantly different than the pdf for AB.  Calculated from all
careers ending in the 90-year
period 1920-2009, these extremely right-skewed distributions for career longevity  extend
over more than three orders of magnitude.
For comparison of the scaling regime, we plot a power-law with exponent $\alpha =1$ (dashed black line). For contrast, we
also plot  normal distributions with identical mean and standard deviation (solid black curve), which on logarithmic
axes, appear to be similar to a uniform distribution over the range $x\in [1, \langle x \rangle+3\sigma_{x}]$. Clearly,
career longevity is not in agreement with a traditional Gaussian ``bell-curve" pdf.  The mean and standard deviation for
IPO data is $\langle x \rangle=  1300 \pm 2100$ and for AB data is $\langle x \rangle= 950 \pm 1800$. }
\end{figure}

\begin{figure*}
\centering{\includegraphics[width=0.84\textwidth]{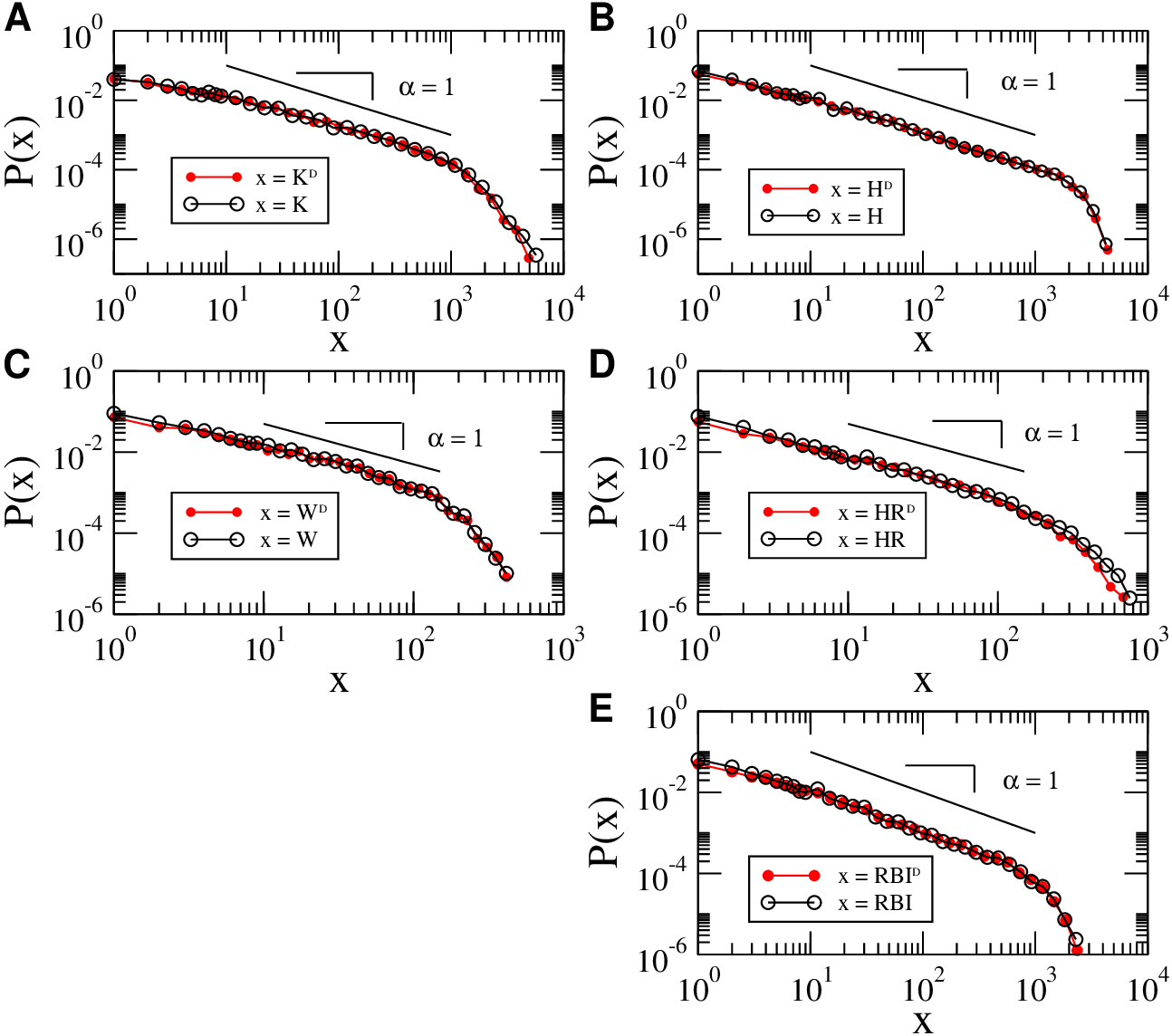}}
 \caption{   \label{metricspdf}  A comparison of the probability density functions of  traditional metrics
and detrended metrics. 
There are small deviations across the entire range between the relative frequency of traditional metrics $X$ (open
circles) and
detrended metrics $X^{D}$ (filled circles)  for {\bf (A)}  strikeouts K, {\bf (B)}  hits H, {\bf (C)} wins W, {\bf (D)}
home runs HR, and {\bf (E)}
runs batted in RBI.  The data is computed using players ending their career in the 90-year period 1920-2009.  The distributions
are invariant under detrending. This
invariance implies that the statistical regularity observed in the right-skewed distributions for longevity and
success in professional sports arises from both the wide range of intrinsic talent among athletes and the underlying 
nature of competition, and not time-dependent factors which may artificially inflate relative success rates.  These extremely right-skewed pdfs for career metrics
extend over more than three orders of magnitude.
For visual comparison, we show a power-law with exponent $\alpha =1$ (straight black lines). 
 }
\end{figure*}

This paper is organized as follows: in Section \ref{sec:Longevity} we first analyze the distribution of career longevity
and success for all players in Sean Lahman's Baseball Archive \cite{BBStat}, which has player data for the 139-year
period 1871-2009.   We plot in Fig. \ref{long} and Fig. \ref{metricspdf}  the
probability density function (pdf) of career longevity and success for several statistical categories. We use this common graphical method to illustrate
the range and frequency of values that historically occur and to uncover information about the complexity of the
underlying system. In Section \ref{sec:Detrend} we motivate a detrending method, and provide examples from other fields.
In Section \ref{sec:Method} we quantify the mathematical averages used to remove the league-wide ability trends that are  time-dependent. 
In Section \ref{sec:Results} we discuss both the surprising and the intuitive results of detrending, with some examples
of careers that either ``rose" or ``fell" relative to their traditional rank. In the electronic-only supplementary information (SI) section we present 10 tables listing the Top-50
All-Time ranking of player accomplishments (both career and seasonal) for traditional metrics versus detrended metrics.
In Section \ref{sec:benchmarks} we use the pdf for each statistical category to quantify statistical benchmarks that
distinguish elite careers, for both traditional and detrended metrics.

\subsection{Career Longevity} 
\label{sec:Longevity} 
The recent availability of large  datasets, coupled with unprecedented computational power, has resulted in many large
scale studies of human phenomena that would have been impossible to perform at any previous point in history
\cite{CompSocScience}. 
A relevant study \cite{BB1} analyzes the surprising features of  career length in Major League Baseball from the
perspective of statistical physics and first uncovered the incredibly large disparity between  the numbers of ``one-hit
wonders" --- those players who one has probably never heard of and probably never seen --- and the ``iron horses" ---
those
legends who lasted at the upper tier of professional baseball for several decades and who became household names.  Using
methods from statistical physics, a recent study \cite{BB2} quantitatively analyzes the {\it rich-get-richer} ``Matthew
Effect" \cite{Matthew} and demonstrates that the universal distribution of career longevity, which is empirically
observed in academics and for several  professional sports, can be explained by a simple model for career progress.

Career length is the most important statistical quantity for measures of career success in professional sports.  This is
because all success measures, such as home runs or strikeouts, are obtained in proportion to the number of opportunities
that constitute the career length. Furthermore, a player's number of successes is constant in time upon retirement, as opposed
to e.g. scientific citations or musical record sales which continue to grow after career termination. In baseball, in
the simplest sense, an opportunity is a plate appearance (for batters) or a mound appearance (for pitchers). 

In this paper, we only consider the two following metrics for player $opportunity$: 
\begin{itemize}
\item[(i)] a player's number of at-bats $(AB)$ and
\item[(ii)] a pitcher's innings pitched in outs $(IPO)$. 
\end{itemize}
These definitions follow from the unique style of baseball, which is simultaneously a single-opponent game (pitcher versus batter) and a team game (offense versus defense). In each ``iteration'' of the game of baseball, a batter faces a pitcher, and the outcome of this contest is either an ``out'' or the advancement of the batter onto the bases. The metrics IPO and AB quantitatively account for these two types of outcomes.
Although other definitions for opportunity can be justified, e.g. plate 
appearances (PA), the salient results are not sensitive to the exact definitions. 

We  analyze data from Sean Lahman's
Baseball Archive \cite{BBStat} which has batting and pitching data for over 17,000 players. In this paper, we do not
distinguish between pitchers and fielders in the case of batting statistics, which neglects the appearance variations
arising from the designated hitter (DH)  rule  in the American League, whereby a substitute is allowed to bat for the pitcher during the game \cite{justify}. 


To compare players across all eras, with and without detrending, we utilize the graphical representation of the
probability density function (pdf).  This is a first step to understanding the frequency of particular types of careers.

The power law pdf
 \begin{equation}
 P(x)\sim x^{-\alpha} \ ,
 \end{equation}
 is found in empirical studies of many complex systems where competition drives the
dynamics, with examples ranging from blockbuster Hollywood movies to human sexuality 
\cite{wealth,citations,Mantegna,Albert,sex,musicians,Hollywood}. An important feature of the scale-free power law is
the large disparity between the most probable value and the mean value of the distribution \cite{MEJN,ClausetPowLaw},
where the most probable value  $x_{mp} \sim 1$, while the mean value $\left<x\right> $ is infinite for $\alpha \leq 2$. 
This is in stark contrast to the Gaussian (Normal) distribution pdf,
\begin{equation}
P(x) = \frac{1}{\sqrt{2 \pi \sigma^{2}}} e^{-(x-\langle x \rangle)^{2}/2\sigma^{2}} \ ,
\end{equation}
for which the mean value and the most probable value coincide.
In data analysis, it is often sufficient to study only the
mean and standard deviation of a sample population since, for data corresponding to a Gaussian distribution, the
fluctuations around the mean decrease according to the central limit theorem as $\sigma \sim 1/\sqrt{N}$. However, this
approach is only well-suited for statistical distributions that are centered around a typical mean value $\langle x
\rangle$ with fluctuations that have a typical scale measured by the standard deviation $\sigma$.  

In the case of the career
statistics analyzed here, which are extremely right-skewed, analyzing only $\langle x \rangle$  overlooks the incredibly large range
of values which includes monumental events that occur relatively frequently compared to the predictions of a Gaussian
distribution. 
Indeed, in the cases of extremely right-skewed distributions, a typical scale for career length is not well-defined.

\begin{figure}
\centering{\includegraphics[width=0.45\textwidth]{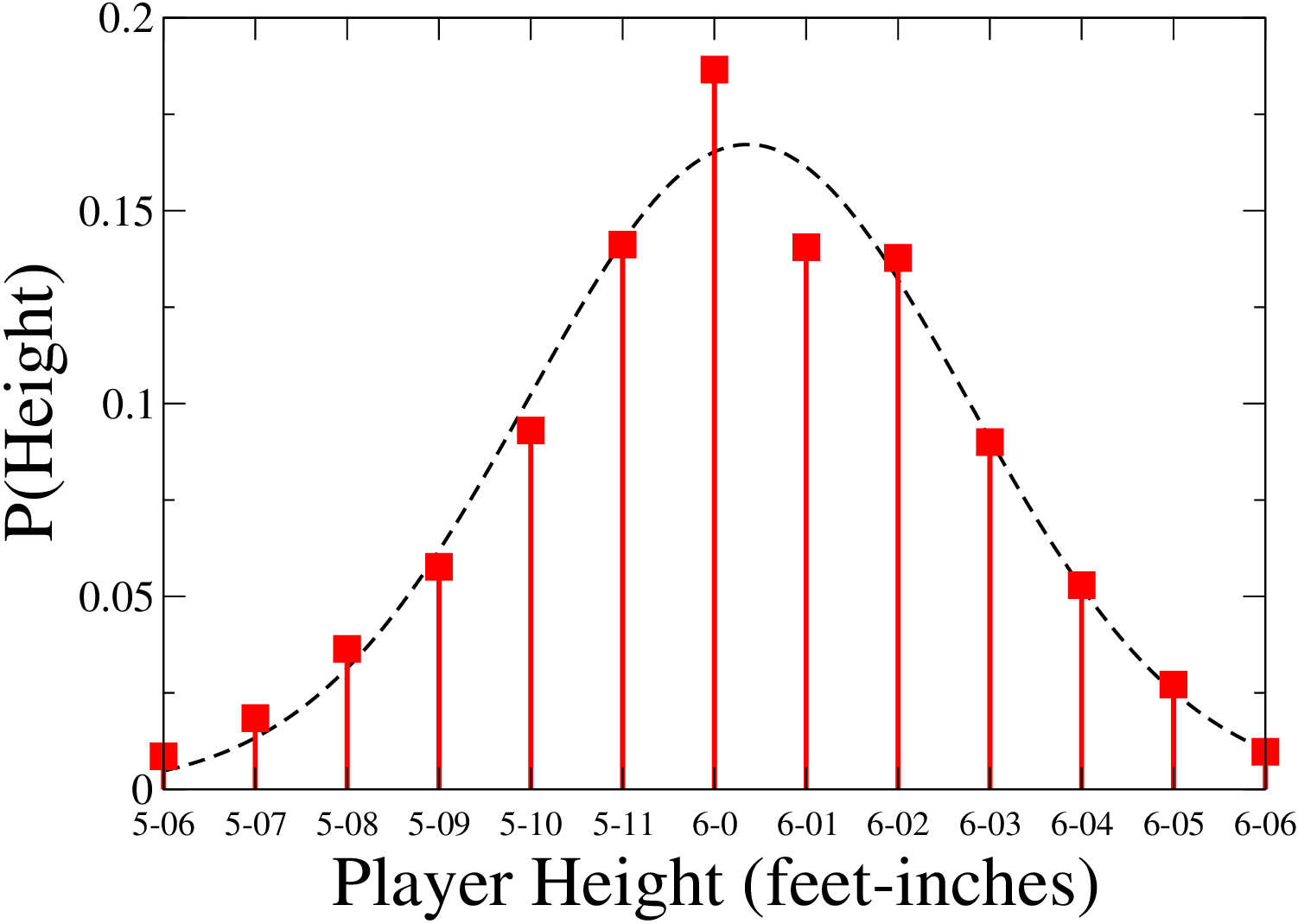}}
  \caption{
  \label{playerheight} A demonstration of a probability density function that has a characteristic scale. The pdf
of Major League Baseball player height.
 The data are fit well by a
 Gaussian ``bell-curve" pdf
(dashed line) with an average height of $6.0$ feet $ \pm 2$ inches.  Data courtesy of baseball-almanac.com, 
accessed at: \newline {\scriptsize http://www.baseball-almanac.com/charts/heights/heights.shtml}}
\end{figure}

In order to emphasize the disparity  between the long and short careers, consider the ratio of the longest career (Pete
Rose, $14,053$ at-bats) to the shortest career (many individuals with one at-bat), which is roughly $10,000$.  For
comparison, the ratio of the tallest baseball player (Jon Rauch, $6$ feet $11$ inches) to the shortest baseball player
(Eddie Gaedel,  $3$ feet $7$ inches) is roughly $2$.  The relatively small value of the player height ratio follows from
the properties of the Gaussian distribution, which is well-suited for the description of height in a human population.
For a qualitative description that is analogous to a pdf, see the histogram of player height plotted in
Fig.~\ref{playerheight}, which demonstrates that the typical height of MLB players is 6 feet $\pm$ 2 inches.

The statistical regularity describing power-law behavior with exponent $\alpha \approx 1$ can be  roughly phrased as
such: for every Mickey Mantle ($8,102$ career at-bats), there are roughly $10$ players with careers similar to Doc ``the
Punk" GautreauÕs ($806$ career at-bats); and for every Doc ``the Punk" Gautreau there are roughly  $10$ players with
careers similar to Frank ``the Jelly" Jelincich  ($8$ career at-bats with one hit). This statistical property arises
from the ratio of frequencies 
\begin{equation}
P(x_{1})/P(x_{2})\sim (x_{1}/x_{2})^{-\alpha} \ ,
\end{equation}
which only depends on $\alpha$ and the scale-free ratio $x_{1}/x_{2}$. This  statistical regularity also applies to the
pdfs that quantify career success metrics for hits (H), home runs (HR), runs batted in (RBI), wins (W) and strikeouts (K).

Thus, in power law distributed phenomena, there are rare extreme events that are orders of magnitude greater than the
most common events. For the pdfs of career longevity and  success analyzed in this paper for which $\alpha <1$, we observe truncated power-law
pdfs 
\begin{equation}
 P(x)\sim x^{-\alpha}e^{-(x/x_{c})} \ ,
 \label{Gamma}
 \end{equation}
 which have a finite mean $\langle x \rangle$ and standard deviation $\sigma$.
   The truncated power-law distribution captures the surprisingly wide range of career lengths that emerge as a result
of  the competition for playing time at the top tier of professional sports. The truncation of the scaling regime
results primarily from the finite length of a player's career. 

\subsection{Detrending}
\label{sec:Detrend} Detrending is a common method used to compare observations made at different times, with
applications in a wide range of disciplines such as economics,  finance, sociology,  meteorology, and medicine. 
Detrending with respect to price inflation in economics is commonly referred to as ``deflation" and relies on a consumer
price index (CPI). The CPI allows one to properly compare the cost of a candy bar in 1920 dollars to the cost of a candy
bar in 2010 dollars.  In stock market analysis, one typically detrends intraday volatility by removing the intraday
trading pattern corresponding to relatively high market activity at the beginning and end of the market day, which
results in a daily activity trend that is ``U-shaped".  In meteorology,  trends are typically cyclical, corresponding to
daily, lunar, and annual patterns,  and even super-annual patterns as in the case of the El Nino effect.   Cyclical
trends are also encountered in biological systems, as in the case of protein concentration fluctuations over cell life
cycles. In baseball, the trends that we will analyze are those that are associated with player performance ability, or
prowess.

It is common for  paradigm shifts to change the nature of  business and the patterns of success in competitive
professions. Baseball has many examples of paradigm shifts, since the game has changed radically since its conception over a
century ago.   As a result, the relative value of accomplishments depends on the underlying time period.
For example,  although a home-run will always be a home-run, and a strikeout will always be a strikeout, the rate at
which these two events occur has changed drastically over time. 

 A relevant  historical  example is the case of Babe Ruth. Before Ruth,  home runs were much
less frequent than they are in 2010. However, following changes in the rule set accompanied by Babe Ruth's success in
the 1920's, many sluggers emerged that are summarily remembered for their home run prowess.  The main time-dependent
aspect we consider in this paper is the variation in relative player ability, a generic concept that can be easily
applied to other professions. Ref. \cite{BB1} finds clear evidence for non-stationarity in the seasonal home-run ability, both on the career and the seasonal level. By comparing the pdfs for career home runs for players belonging to either the 1920--1960 or the 1960--2000  periods, it is shown that the pdf for career  home-runs are shifted towards larger totals in the more-recent 1960--2000 period. Moreover, by comparing the pdfs for seasonal home-run ability for players belonging to one of the three periods 1940--1959, 1960--1979, or 1980--2006, it is shown that  at the fundamental seasonal time-scale, the  home-run rates among players is also changing, where the pdf is becoming more right-skewed with time. 
These results show why it is important to account for the era-dependence of statistics  when comparing career statistical totals. 

Yet, this is not the only time-dependent factor that we consider.  By detrending, we remove the net trend resulting from
many underlying factors, season by season, which allows the proper (statistical) comparison of contemporary players to
players of yore (of lore). A significant result of this paper is that detrending for seasonal prowess maintains the
overall  pdf of success while re-ordering the ranking of player achievements locally.  This means that the
emergence of the right-skewed pdfs for longevity and  success are not due to changes  in player ability, but rather,
result from the  fundamental nature of competition.

The idea behind detrending is relatively straightforward. By calculating the  average prowess of all players in a given
season, we effectively renormalize all statistical accomplishments  to the typical prowess of all contemporaneous
competitors.  Hence, detrending establishes relative significance levels, such that  hitting fifty home runs was of less
relative significance during the ``Steroids Era" than hitting fifty home runs during the 1920's. The objective of this
work is to calculate the detrended statistics of a player's whole career. To this end, we compare career metrics that
take into account the time-dependence of league-wide player ability.  While there is much speculation and controversy
surrounding the causes for changes in player ability, we do not address these individually. In essence, we blindly
account for not only the role of PED \cite{Ped,NEJM,BritFB,NYT,Surg,SteroidsTobin,PEDMitchell, DeVany}, but also changes
in the physical construction of bats and balls, sizes of ballparks,  talent dilution of players from expansion \cite{HHH,parity}, etc.

\section{Materials and Methods} 
\label{sec:Method}
\subsection{Data} We analyze historical Major League Baseball (MLB) player data compiled and made publicly available by
Sean Lahman  \cite{BBStat}. 
The Lahman Baseball Database is updated at the end of each year, and has player data dating back to 1871. In total, this
database records 
approximately 35,000 players seasons and approximately 17,000 individual player careers.

\subsection{Quantifying Average Prowess} 
\begin{figure}
\centering{\includegraphics[width=0.45\textwidth]{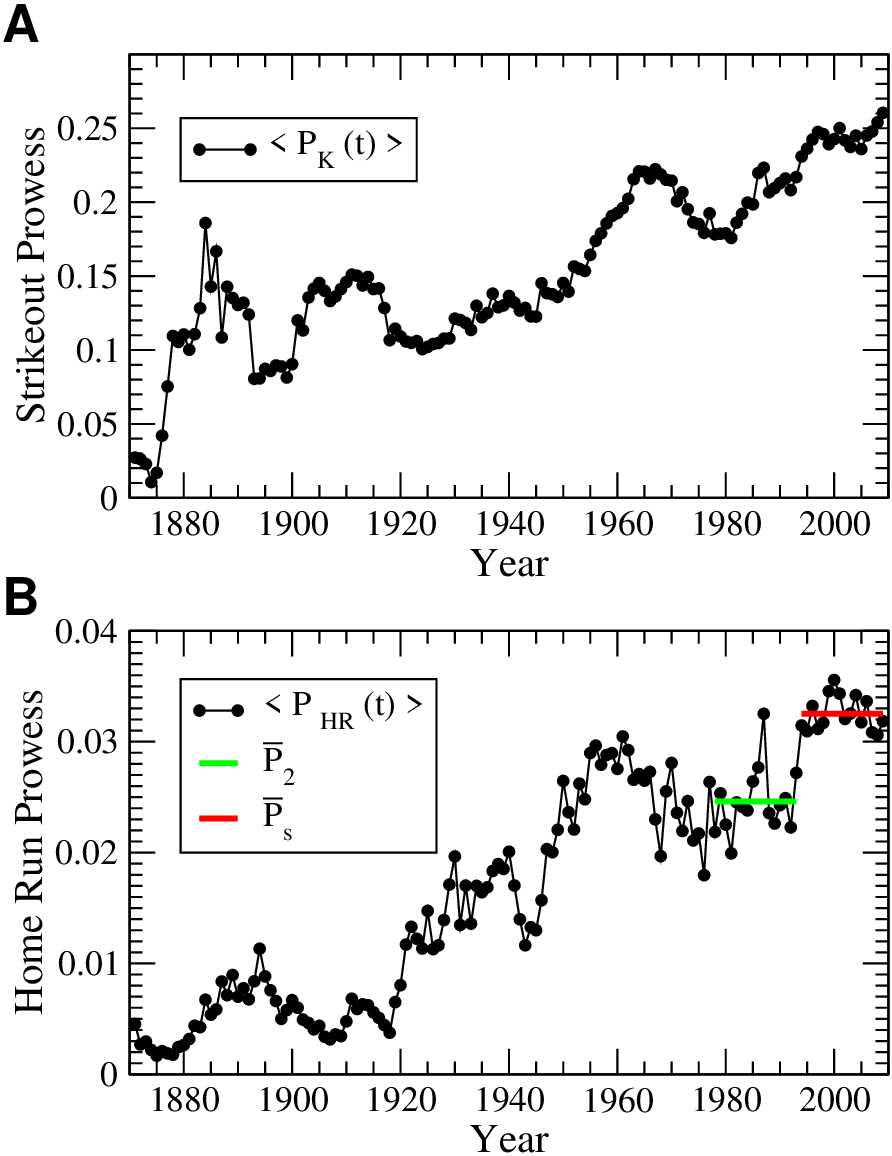}}
  \caption{
  \label{prowesstrend}  Success rates reflect the time-dependent factors that can inflate or
deflate measures of success.
The annual prowess for {\bf (A)} strikeouts (K)
 and {\bf (B)} home runs (HR) are calculated
using Eq. (\ref{prowess}).  Prowess is a weighted measure of average league wide ability, with more active players
having a larger statistical weight than less active players in the calculation of the prowess value  $\langle P
\rangle$. {\bf (B)} We also plot the average values $\overline{P}_{1}$ and $\overline{P}_{s}$ of $\langle P_{HR}(t)
\rangle$ over the 16-year periods $\{Y_{1}\}\equiv 1978-1993$ and $\{Y_{s}\}\equiv 1994-2009$, where the latter period
roughly corresponds to the ``steroids'' era. We calculate $\overline{P}_{1}= 0.025 \pm 0.003$ and  $\overline{P}_{s}=
0.033 \pm 0.002$, and find $\overline{P}_{1} < \overline{P}_{s}$ at the $0.005$ confidence level. For the difference
$\Delta \equiv 
\overline{P}_{s} - \overline{P}_{1}$, we calculate the confidence interval $0.005 < \Delta < 0.010$ at the $0.01$
confidence level. }
\end{figure}

We define prowess as an individual player's ability to achieve a success $x$
(e.g. a home run, strikeout) in any given opportunity $y$ (e.g. an AB or IPO).   In Fig. \ref{prowesstrend} we plot the average annual prowess for strikeouts
(pitchers) and home runs (batters) over the 133-year period 1876-2009 in order to investigate the evolution of player
ability in Major League Baseball.  The average prowess serves as an index for comparing accomplishments in distinct
years.  We conjecture that the changes in the average prowess are related to  league-wide factors which can be quantitatively removed (detrended) by normalizing accomplishments by the average prowess for a given season. 

We first calculate the   prowess $P_{i}(t)$ of  an individual player $i$  as
\begin{equation}
P_{i}(t) \equiv x_{i}(t)/y_{i}(t) \ ,
\end{equation} 
 where $x_{i}(t)$ is an individual's total number of successes out of his/her total number of opportunities $y_{i}(t)$ in a given  year $t$. 
 To compute the  league-wide average prowess, we then compute the weighted average for season $t$ over all players 
\begin{equation}
 \langle P(t) \rangle \equiv \frac{\sum_{i} x_{i}(t)}{\sum_{i} y_{i}(t)} = \sum_{i} w_{i}(t)P_{i}(t) \ ,
 \label{prowess}
 \end{equation}
 where
 \begin{equation}
  w_{i}(t) = \frac{y_{i}(t)}{\sum_{i}y_{i}(t)} \ .
  \end{equation}
The index $i$ runs over all players with at least  $y'$ opportunities during year $t$, and $\sum_{i}y_{i}$ is the total
number of opportunities of  all $N(t)$ players during  year $t$. We use a cutoff $y' \equiv 100$ which eliminates
statistical 
fluctuations that arise from players with very short seasons.

We now introduce the detrended metric  for the accomplishment of player $i$ in year $t$,
 \begin{equation}
 x^{D}_{i}(t) \equiv x_{i}(t) \ \frac  {\overline{P}}{\langle P(t) \rangle} \,
 \label{xdsingle}
  \end{equation}
  where $\overline{P}$ is the average of  $\langle P(t) \rangle$ over the entire period, 

 \begin{equation}
 \overline{P} \equiv \frac{1}{110}\sum_{t=1900}^{2009}\langle P(t) \rangle \ .
  \end{equation}
 The choice of normalizing with respect to $\overline{P}$ is arbitrary, and we could just as well normalize with respect
to $P(2000)$, placing all values in terms of current ``2000 US dollars," as is typically done in economics.

\begin{figure*}
\centering{\includegraphics[width=0.75\textwidth]{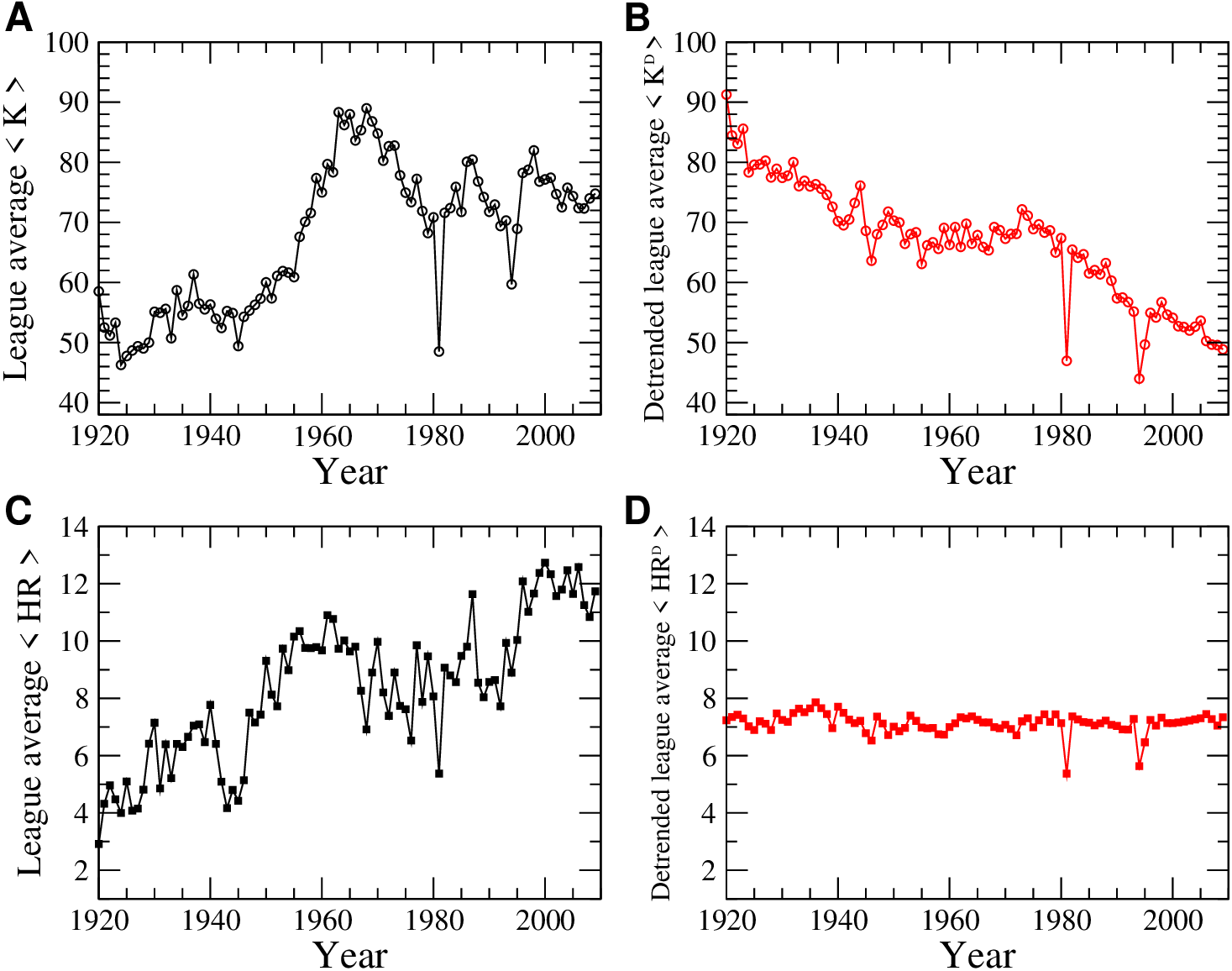}} 
  \caption{
  \label{detrendedave}  A comparison of traditional and detrended league averages demonstrates the utility of
the detrending method. Annual
per-player averages for {\bf (A)} strikeouts {\bf (B)} detrended strikeouts (of pitchers),
 {\bf (C)} home run, and {\bf (D)} detrended home runs (of batters). The detrended home run average is remarkably constant over the
90-year ``modern era" period
1920-2009, however there remains a negative trend in the detrended strikeout average. This residual trend in the
strikeout average may result from
the decreasing role of starters (resulting in shorter stints) and the increased role in the bullpen relievers, which affects the average number of opportunities obtained for players in a given season. This follows from the definition of the detrended average given by Eq. (\ref{xDave}).
 A second detrending for average innings pitched per game might remove this residual trend demonstrated in Fig. \ref{aveIPOpG}. The sharp negative
fluctuations in 1981 and 1994-1995 correspond to player strikes resulting in season stoppage and a reduced average number of opportunities $\langle y(t) \rangle$  for these seasons.}
\end{figure*}

 In Fig. \ref{detrendedave} we compare the seasonal average of $\langle x(t) \rangle$ to the prowess-weighted average
$\langle x^{D}(t) \rangle $, for  strikeouts per player  and home runs per player. We define $\langle x(t) \rangle$ as
 \begin{eqnarray}
 \langle x(t) \rangle &=& \frac{1}{N(t)}\sum_{i}x_{i}(t)  \nonumber \\
 &=& \langle P(t) \rangle \frac{\sum_{i} y_{i}(t)}{N(t)}  = \langle P(t) \rangle  \langle y(t) \rangle  
 \label{xave}
  \end{eqnarray}
and $\langle x^{D}(t) \rangle $ as,
\begin{eqnarray}
\langle x^{D}(t) \rangle &=& \frac{1}{N(t)}\sum_{i}x^{D}_{i}(t) = \frac{ \overline{P}}{\langle P(t) \rangle N(t)} \sum_{i}x_{i}(t) \nonumber \\
 &=& \overline{P} \  \frac{ \langle x(t) \rangle}{\langle P(t) \rangle }   = \overline P \ \langle y(t)\rangle \ .
 \label{xDave}
  \end{eqnarray}
 As a result of our detrending method defined by Eq. (\ref{xdsingle}), which removes the time-dependent factors that affect  league-wide ability  from the average number of successes across all players in a given season, we find that Eq. (10) is independent of $\langle P(t) \rangle.$ 
The averages computed in Eq. (\ref{xave}) and Eq. (\ref{xDave}) are computed for players with $y_{i}(t)>y' \equiv 100$ 
appearances and $x_{i}(t) > x' \equiv 0$ successes, to eliminate statistical fluctuations arising from insignificant
players. 

Using this method, we calculate detrended metrics for baseball players for both single season ($x_{i}^{D}$ corresponding
to Eq.~(\ref{xdsingle})) and total career accomplishments ($X_{i}^{D}$ corresponding to Eq.~(\ref{xdtotal})). 
Naturally, the detrended career metric is  cumulative, which we calculate for each player $i$ over his career as 
\begin{equation}
X^{D}_{i} \equiv \sum _{s=1}^{L} x^{D}_{i}(s) \ , 
\label{xdtotal}
\end{equation}
where $s$ is the season index and $L$ is the player's career length measured in seasons.

\section{Results}
\label{sec:Results}
\subsection{Comparing across historical eras} In this section we  discuss the results of detrending. We begin with the
analysis of career longevity, which we discuss briefly, and refer interested readers to \cite{BB1} for a more detailed
discussion. The main take-home result from Fig.~\ref{long} is the variety of  career length of major league players. 
Surprisingly,  we find  that about 3\% of fielding batters (non-pitchers) have their premier and finale in the same AB.
Furthermore, approximately 5\% of all fielding batters finish their career with only one hit. Similarly,  3\% of all
pitchers complete their career with an inning or less of pitching. Yet, remarkably, there are several players with
careers that span more than 2,000 games, 10,000 at bats, and 4,000 innings. This incredible range is captured by the 
pdfs of career longevity which appear as linear when plotted on log-log scale.

The statistical regularity of the pdfs in Fig. \ref{long} and Fig. \ref{metricspdf} allows for the quantitative
comparison of careers. Furthermore, this statistical regularity is {\it very} different from the statistical regularity
captured by the common Gaussian  (Normal) ``bell curve". For comparison, we plot Gaussian pdfs in
Fig.~\ref{long}  which have the same mean and standard deviation as the data represented by the power law pdf. On a
log-log scale, the corresponding Gaussian pdfs appear as uniformly distributed up until a sharp cutoff around $10^{4}$
appearances. The striking disagreement between the data and the  Gaussian bell-curves is evidence of the complexity
underlying career longevity. The truncated power-law, defined in Eq.~(\ref{Gamma}), has two parameters, $\alpha$ and
$x_{c}$, where we observe values $\alpha \lesssim 1$ for all metrics studied. The value of $x_{c}$ marks the beginning of the
exponential cutoff which separates the exceptional careers for which $x>x_{c}$ from the more common careers
corresponding to $x <  x_{c}$.

In Figs.~\ref{metricspdf}.A- \ref{metricspdf}.E we plot the pdfs for several statistical categories, and for each we
plot both the traditional and detrended metrics.  We observe only a slight variation between the
traditional and detrended pdfs, on a log-log scale.  This is an indication that the detrending method only makes
relatively small refinements  to the career totals. For a given player, the ratio between his detrended and traditional
metrics $r \equiv X^{D}/X $ is closely distributed around a value of unity for all metrics analyzed (see Table
\ref{table:stats}). 
Interestingly, for the case of career home runs, the distribution of $r$-values across all players is bimodal, with
$\langle r \rangle = 1.2 \pm 1.0$, which accounts for the the slight deviation in the tail of the pdf for detrended home
runs in  Fig.~\ref{metricspdf}.D. 

\begin{figure}
\centering{\includegraphics[width=0.45\textwidth]{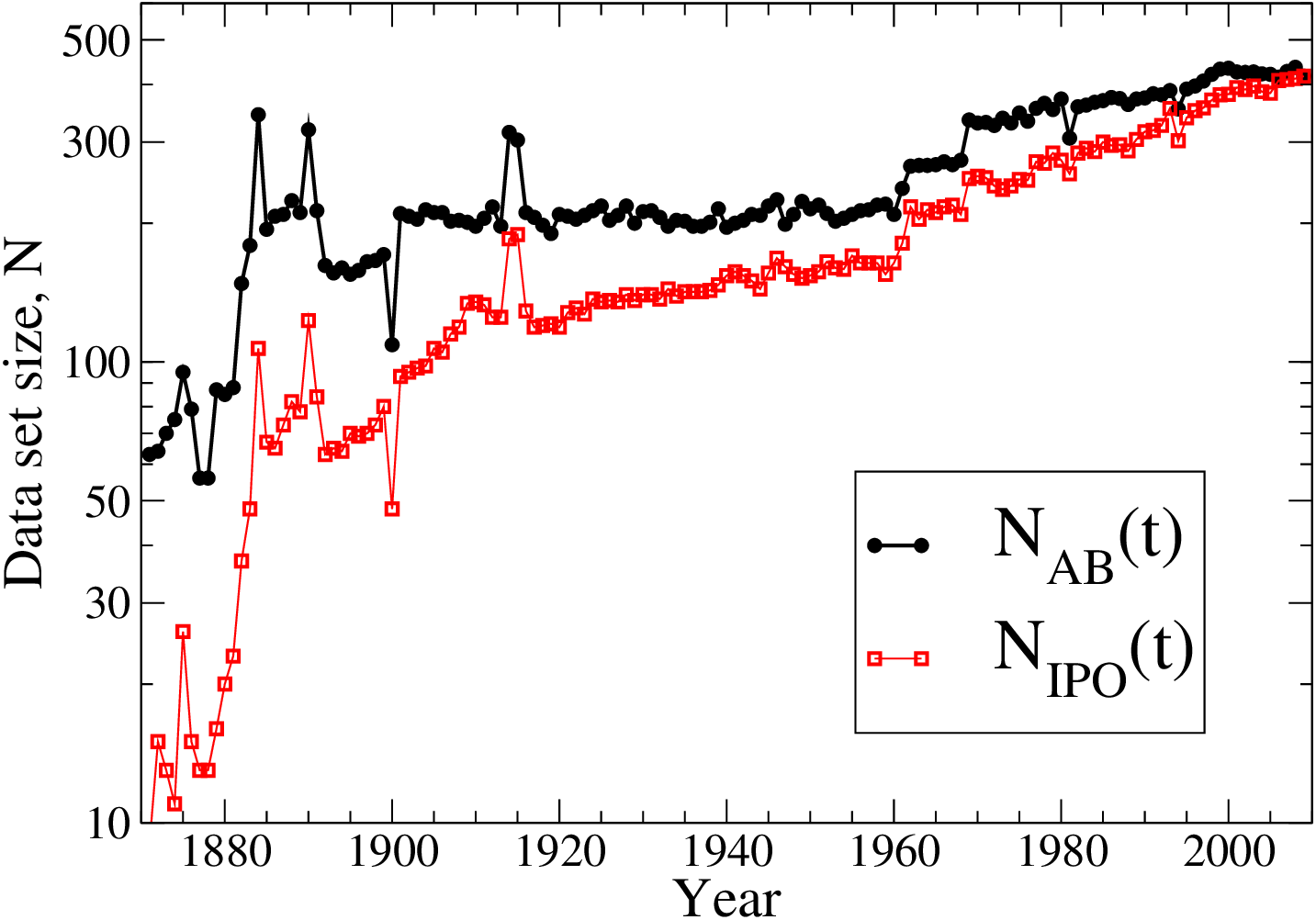}}
  \caption{
  \label{Nplayers} Semi-log plot of data set size exhibits the  growth of professional baseball. The
size of
the data sets, $N$, is used to compute the annual trends for pitchers and batters in Fig. \ref{prowesstrend} and Fig.
\ref{detrendedave}. These data sets correspond to cutoffs $x' \equiv 0$ and $y' \equiv 100$ in Eqs.
(\ref{prowess})-(\ref{xDave}).  Interestingly, we observe a spike in league size during $WWI$, possibly corresponding to
the widespread replacement of league veterans with multiple replacements through the course of the season.  We also note
that one can clearly see the jumps associated with the expansion between 1960 and 1980. }
\end{figure}

In Fig.~\ref{prowesstrend} we plot the trends in average prowess $\langle P(t) \rangle$, defined in Eq.~(\ref{prowess}),
for pitchers obtaining strikeouts and batters hitting home runs over the years 1876-2009. The number of players $N(y')$
entering into the calculation of each data point depends on the cutoff $y'$ and varies according to the completeness of
the database and to the number of players  on MLB rosters (see Fig. \ref{Nplayers}).  In the case of both  home runs and
strikeout ability, there is a marked increase with time reflecting the general changes in game play, player ability, and
the  competitive advantage between pitcher and batter. For pitchers, there are two significant periods corresponding to
the dead-ball era (approximately 1900-1920) and to the era around the 1960's when pitchers were relatively dominant
(ending when the mound was officially lowered in 1969 by league wide mandate). The period beginning in the 1920's shows
a clear increase in home run ability, often attributed to the outlawing of the spitball and the emergence of popular
sluggers, such as Babe Ruth and Rogers Hornsby.  Together, Fig.~\ref{prowesstrend}.A and Figs.~\ref{prowesstrend}.B 
indicate that the average prowess  of major league players is non-stationary. 

\begin{figure}
\centering{\includegraphics[width=0.48\textwidth]{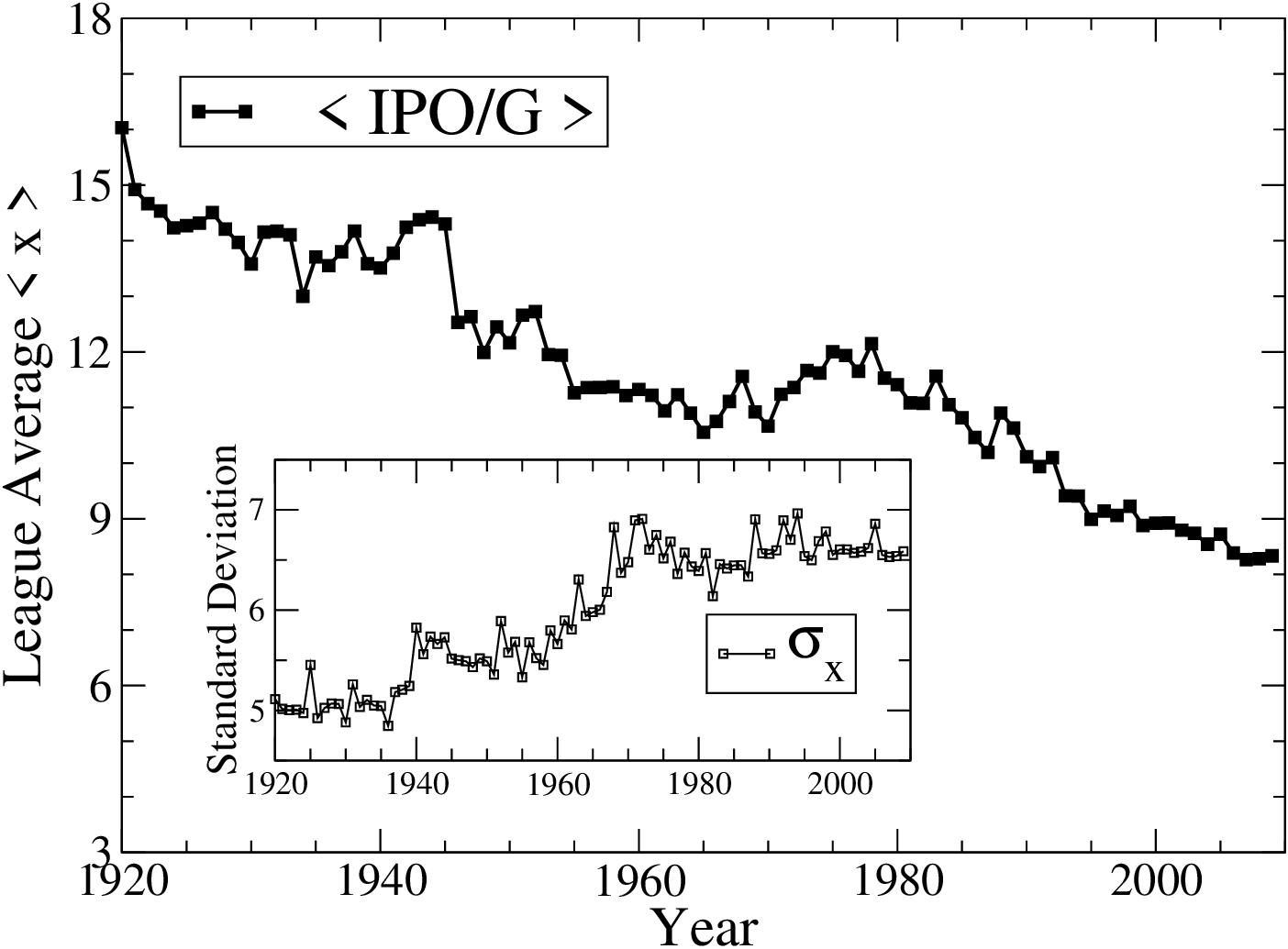}}
  \caption{
  \label{aveIPOpG}  Annual per-player average for innings pitched per game appearances (IPO/G)
demonstrates strategy change in  pitcher use.
 A clear
decreasing trend in the average number of innings pitched per outing demonstrates the increased role of relief pitchers.
 This is a strategic trend, which we do not address in this paper, as we have restricted our analysis to player ability
trends. This explains why detrending for pitching ability (see Fig. \ref{detrendedave}) does not yield a constant
average over time (as in the case with detrended  home-run ability).  We use a cutoff $y'=10$ games in computing
this average. }
\end{figure}

Closely related, but not equal to average prowess $\langle P(t) \rangle$, is the average number of successes $\langle x
\rangle$ (eg. home runs, hits, strikeouts, wins, etc) per player for a particular season. Comparing Eq.~(\ref{xave}) and
Eq.~(\ref{xDave}), the resulting relationship $\langle x^{D}(t) \rangle \equiv  \langle y(t) \rangle \overline{P}$ may
appear trivial, since the quantity $\langle x^{D}(t) \rangle$ does not depend on yearly player abilities, but rather
yearly player opportunities $\langle y(t) \rangle$. This is entirely the  motivation for detrending: we aim to remove
all factors contributing to league-wide player ability.  Thus by effectively removing the
seasonal trends from each player's metrics we measure accomplishments based only on comparable opportunities.  We
graphically illustrate this result in Fig.~\ref{detrendedave} where we plot the average number $\langle x \rangle$ of
strikeouts and home runs
 next to the detrended average number   $\langle x^{D} \rangle$ of strikeouts and home runs, per pitcher and batter,
respectively, over the years 1920-2009. 

We also observe in
Fig.~\ref{detrendedave}.C an approximately linear decrease in the detrended league average for strikeouts  over time.
This is a result of the increased use of relief pitchers (pitchers who do not regularly start games, but rather replace the starting pitchers on demand) in baseball, which reduces the
number of innings pitched per pitcher.  This, however, is a strategic change in the way the game of baseball is managed,
and does not reflect player ability.  In order to detrend for this secondary effect, which might make possible a
comparison between Cy Young, Greg Maddux and Mariano Rivera, one must further detrend pitching statistics by the league
average of innings per game, which we illustrate in Fig. \ref{aveIPOpG}. The increase in the standard deviation of IPO/G
over time is due to the increased variability in the type and use of pitchers in baseball.

The relatively constant value for  the detrended league average of home-runs in Fig. \ref{detrendedave}.D demonstrates
that this method properly normalizes home-run statistics across time.  As a result, the baseline  for seasonal
comparison is approximately 7 home runs per season over the entire 90-year period 1920-2009. We use this relatively constant baseline to correctly compare the single-season accomplishments  of Babe Ruth, in each of his historical seasons 1920-1933, with
the single-season accomplishments of Roger Maris' 61 in '61, with the wild race between Mark McGwire and Sammy Sosa in 1998, and with Barry Bonds' pinnacle performance
in 2001. In Table
VII we rank the top-50 individual home run performances by season, both for the traditional and detrended metrics.
Overwhelmingly, Babe Ruth's accomplishments in the 1920's are superior to those of his peers. Surprisingly, there are
no modern players later than 1950 that make the top-50 list, not even the career years of George Foster ($x^{D}=40$ in
1977), Barry Bonds ($x^{D} = 42$ in 2001), and Mark McGwire ($x^{D}= 44$ in `98). Even the 61 home runs of Roger Maris in
1961 is
discounted to $x^{D} = 40$ detrended home runs, placing him in a tie for $97^{th}$ on the detrended ranking list. 

 In fact, in 1961, the season was 
8 games longer than in 1927, when Babe Ruth hit a seemingly insurmountable 60 home runs. There was much public resentment  over Babe Ruth's seminal record being broken, which 
caused the commissioner of baseball in 1961 to suggest placing 
an asterisk next to Roger Maris' record \cite{KB99}; In analogy,  it is  being suggested that the  asterisk be used to denote the conditional achievement of baseball players
found guilty of using PED. Interestingly, after the 1961 season, the commissioner reacted to the sudden increase in home runs  by expanding the strike zone by league mandate, which resulted in the 
competitive balance tipping in favor of the pitchers during the 1960's. On several historical occasions, the competitive balance was shifted by 
such rule changes. Nevertheless, the detrending method accounts for variations in  both season length and various performance factors, since this method 
normalizes achievement according to local seasonal averages. 
 The 
results of detrending may be surprising to some, and potentially disenchanting to others.  We note that these results {\it do not} mean
that Babe Ruth was a better at hitting home runs than any other before or after him, but rather, {\it relative} to the
players during his era, he was the best home run hitter, by far. 

 Table 2 lists the top-50 career home-run rankings,  and Tables 3-11 list the top-50 rankings for several other statistical categories. In each table, we compare the traditional rankings alongside
the detrended rankings. Tables 2-4  list the career batting statistics for HR,  H and RBI,
 while Tables 5 and
6  list career pitching statistics for strikeouts K and wins W. We choose these
statistics because four of these categories have a popular benchmark associated  with elite careers, the 500 home run,
3000 hit, 3000 strikeout and 300 win ``clubs''. 

Each table provides two parallel rankings, the traditional rank and the detrended rank.  The column presenting
the traditional rank lists  from left to right the traditional ranking, the player's name, the player's final season,
with the total number of seasons played (L), and the  career total $X$. The column presenting the
detrended rank lists from left to right the detrended ranking $Rank^{*}$ with the corresponding traditional rank
$(Rank)$ in parenthesis, the relative percent change in the rankings $\% Change = (Rank^{*}-Rank)/Rank$, the player's
name, the player's final season with the total number of seasons  (L), and the detrended career
total $X^{D}$ corresponding to Eq. (\ref{xdtotal}). The tables listing seasonal records have an analogous format, except
that the year listed is the year of the achievement with the number of seasons  into his career ($Y\#$).

In the case of career home runs (see Table 2), Babe Ruth towers over all other players with almost twice as many detrended career home
runs as his contemporary, the third-ranked Lou Gehrig. Hank Aaron's detrended career total is discounted by a
significant factor and he falls to 5th place. Overall, the eras become well-mixed for detrended home runs, whereas the
traditional top-50 is dominated by players from the past thirty years.  Possibly the most significant riser in the
detrended homer run rankings is Cy Williams, who played in the era prior to Babe Ruth.  Cy Williams moves up largely due
to the relatively low home run prowess of the early 1900's.  Nevertheless, according to our statistical analysis, Cy
Williams, along with Gavvy Cravath, are strong candidates for retroactive induction into the Hall of Fame. A similar
amount of mixing occurs for the career strikeouts rankings. Of particular note are Amos Rusie and Dazzy Vance, titans
from a distant era whose accomplishments rise in value relative to the contemporary household names of Nolan Ryan, Randy
Johnson, and Roger Clemens.

In contrast to the vast reordering of the top 50 for home runs and strikeouts, the detrended rankings for career hits
undergoes much less change.  This indicates that hitting prowess has been relatively stable over the entire history of
baseball.  In light of this fact, the batting average (BA), which is approximately a hit per opportunity average, is
likely an unbiased and representative metric for a player, 
even without detrending  time-dependent factors.

The detrended rankings for career wins show the least amount of change of the $5$ statistical categories we studied. We
attribute this stability mostly to the nature of a pitcher earning a win, which is significantly related to the overall
team strength \cite{parity,RednerTeams}.  Since a win results from a combination of factors that are not entirely
attributed to the prowess of the pitcher, detrending does not change the relative value of wins in a player-to-player
comparison.  Wins are also biased towards starting pitchers who, on average, pitch more innings than relievers. The
traditional cumulative metrics for pitching undervalue the accomplishments and role of relief pitchers, that have been
an increasingly important feature of the game as illustrated in Fig. \ref{aveIPOpG}. In order to incorporate middle and
late relief pitchers into an all-inclusive comparison of pitchers, a metric such as strikeouts per inning or earned run
average (ERA) will be more descriptive than career strikeouts or wins. To this end, strikeouts per inning and ERA are
fundamentally averages, and would fall into a separate class of pdf than the truncated power-laws observed for the
metrics considered in this paper.

 
 In Tables 6-11 we list the traditional and detrended single season statistics for HR, H, RBI,
and K. There are too many players to discuss individually, so we mention a few interesting observations. 
First, the rankings for detrended home runs $HR^{D}$ are dominated by seasons prior to 1950.  Second, of contemporary
note, Ichiro Suzuki's single season hits record in 2004, which broke the 83-year record held by George Sisler, holds its
place
as the top  single-season hitting performance of all time.  Finally, 
we provide two top-50 tables for single season
strikeouts in Tables 10 and 11.  We provide two separate tables because the
relative performance of pitchers from the 1800's far surpasses the relative performance of contemporary legends. Hence,
in Table 10 we rank all single-season performances from 1883-2009, while in Table 11 we rank all single-season
performances from the ``modern'' era 1920-present. We note that the dominance of Table 10 by 19th century baseball
players could reflect fluctuations from  small data set size and  incomplete records of pitchers in the 19th century
(see Fig.
\ref{Nplayers}). Still, Matt Kilroy's 513 strikeouts in 1889 seems unfathomable by today's standards, and the seemingly
out of place number may reflect factors which detrending can not account for, e.g. the level of competition being
significantly reduced, as baseball was not a full-time profession for many players in the 19th century. Table 11,
filled with names that are much more familiar,  better illustrates the relative merits of forgotten names such as
Dazzy Vance and Bob Feller. While the relative changes are mostly positive, with Nolan Ryan's six monumental seasons
still notable, there is an unexpected discount of Sandy Koufax's 382 strikeouts in 1965.

\subsection{Calculating Career Benchmarks}
\label{sec:benchmarks}

In this section we outline an approach for calculating a set of statistical criterion that can be used to objectively
define extraordinary careers.  Again, we use historical examples from baseball to illustrate the utility of 
quantifying benchmarks that can be used for distinguishing outstanding performance for professional reward, 
such as annual bonus, salary increase, tenure.
 
Due to the rarity of careers surpassing the benchmarks of 500 home runs, 3000 hits, 3000 strikeouts, and 300 wins, these
milestones are usually accepted by baseball fans and historians as clear indicators of an extraordinary career.
However, these benchmarks are fundamentally arbitrary, and their continued acceptance can probably be attributed to their
popularity with the media personalities that cover baseball. 
 Using the properties of the pdf, that we have shown accurately characterizes many baseball statistics, we can extract
 more objective benchmarks as we now discuss.

We approximate the pdf $P(x)$ of each success metric, $x$, by the gamma distribution,
\begin{eqnarray}
P(x) dx \approx {\rm Gamma}(x; \alpha, x_{c}) dx = \nonumber \\
\frac{(x/x_{c})^{-\alpha} e^{-x/x_{c}}}{ \Gamma(1-\alpha)}
\frac{dx}{x_{c}} \propto x^{-\alpha} \ e^{-x/x_{c}} \ ,
\end{eqnarray}
and use the  mathematical properties of this function in order to define a statistically significant benchmark
$x^{*}$. We calculate the value of $x^{*}$  by using the integral properties of the Gamma distribution.  For a threshold
level $f$, we determine the value of $x^{*}$ such that only $f$ percent of players exceed the benchmark value $x^{*}$,
\begin{eqnarray}
f &=& \int_{x^{*}}^{\infty} \frac{x^{-\alpha}e^{-x/x_{c}}}{x_{c}^{1-\alpha}\Gamma(1-\alpha)}dx \nonumber \\
&=&\frac{\Gamma[1-\alpha,\frac{x^{*}}{x_{c}}]}{\Gamma(1-\alpha)} = Q[1-\alpha, \frac{x^{*}}{x_{c}}] \ ,
\end{eqnarray}
where $\Gamma[1-\alpha,\frac{x^{*}}{x_{c}}]$ is the incomplete gamma function and $Q[1-\alpha, \frac{x^{*}}{x_{c}}]$ is
the regularized gamma function. The regularized gamma function is numerically invertible.  Exploiting this property, we
calculate 
\begin{equation}
x^{*}=x_{c} Q^{-1}[1-\alpha,f] \ ,
\end{equation}
using the inverse regularized gamma function found in standard computing packages. In addition to the analysis performed
in
\cite{BB2}, where a graphical method is used to determine the values $\alpha$ and $x_{c}$ from the pdf using a graphical
least-squares routine, here we use the Maximum Likelihood Estimator (MLE) in order to determine $\alpha$ and $x_{c}$
from the observed values, $x_{i}$, of the entire data set \cite{MLEgamma}. The values of $\alpha$ and $x_{c}$ that
maximize the log-likelihood of observing the data set $\{x\}$ of size $N$ are, to first approximation,
\begin{equation}
\alpha_{MLE} = 1- \frac{3-z + \sqrt{(z-3)^{2}+24 z}}{12 z} \ ,
\end{equation}
where $z$ is calculated from the $N$ individual data values $x_{i}$,
\begin{equation}
z = \ln(\sum_{i=1}^{N}x_{i})- (\sum_{i=1}^{N} \ln x_{i}) / N \ .
\end{equation}
Since the average value of the gamma distribution is $\langle x \rangle = (1-\alpha) x_{c}$, the MLE for
$x_{c}$ is computed using the estimated value of $\alpha$,
\begin{equation}
x_{c}= \frac{\langle x \rangle}{1-\alpha} = \frac{1}{N(1-\alpha_{MLE})}\sum_{i=1}^{N}x_{i} \ .
\end{equation}
Table \ref{table:stats} lists the values of $\alpha_{MLE}$, $x_{c}$, and the corresponding standard deviation
$\sigma_{\Gamma}= x_{c} \sqrt{1-\alpha}$ calculated for the Gamma pdf.  We use a threshold of 2\%, i.e. $f= 0.02$, which
corresponds roughly to the percentage of all baseball players elected into the Cooperstown Baseball Hall of Fame
\cite{BB2}, and then we find the benchmark value can be approximated as $x^{*} \approx 4 \ \sigma_{\Gamma} = 4 \ x_{c} \
\sqrt{1-\alpha}$.
 This approximation is a consequence of the universal scaling form of the gamma function ${\rm Gamma}(x; \alpha,
x_{c}) = {\rm Gamma}(x/x_{c}; \alpha)$,
such that for a given $f$ and $\alpha$, the ratio 
\begin{equation}
\frac{x^{*}}{\sigma_{\Gamma}} = \frac{Q^{-1}[1-\alpha,f]} {\sqrt{1-\alpha}}
\end{equation}
is independent of $x_{c}$. Furthermore, this approximation is  valid for all MLB statistics  since $\alpha$ is
approximately the same for all pdfs analyzed. In Table \ref{table:stats} we compute $x^{*}$ for both traditional $X$
metrics
and detrended $X^{D}$ metrics at the two thresholds $f=0.02$ and $f=0.01$.  We see that the values of $\alpha_{MLE}$ are
approximately equal for both traditional and detrended data sets, and that the values of $x^{*}$ do not vary
significantly. As a result (at the $f=0.02$ level), a player with either $ X=1900$  hits or $X^{D}=1900$ detrended hits is statistically stellar
 in comparison to other players.

\section{Discussion}

The statistical physics of social phenomena is a growing field which aims to describe the complex behavior that emerges
from the  interactions of agents \cite{Mantegna,internetdating,Econophys4,PAsex,mobility,SocialDynamics}.  While it
is often difficult to  account for the interaction complexity in explicit governing equations, a first step towards
understanding the underlying social mechanism
 is to study the macroscopic behavior of the system. The quantitative analysis  of human achievement  in competitive arenas, e.g. sports and academia,  is an open  
  topic of investigation, which in recent studies has  combined methods from sociology, economics, and statistical physics \cite{BB1,Scientists,H0,Cr}. 
   
 Here we analyze the distribution of success  in a population of
competitive athletes.  Because professional baseball has such a standard and precise method for recording historical
achievement, the box-score, we are able to compare the achievements of professional baseball players for over 100 years
of MLB.  In order to account for changes in relative player ability over time, we have developed a detrending method
which accounts for inflationary and deflationary factors and allows for an objective comparison of players across time.
 Remarkably, we find using our detrending method, that the distributions of career success are invariant after we
normalize accomplishments to local
averages. Thus, even controlling for time dependent factors, the distribution of career achievement is extremely right-skewed, and can be quantified by a truncated power-law. Furthermore, in order to distinguish stellar careers, we derive non-arbitrary statistical benchmarks based on
significance thresholds defined by the pdfs of success for the entire population. 

Typically, only the greatest career achievements are recorded in the annals of history. Here we analyze all participants
in a competitive system to compare and contrast the various types of careers.
The statistical regularity of the pdfs quantifying metrics for career longevity and  career success also exist for
athletes in several other professional sports leagues  (American basketball, Korean baseball, English
football, ATP Tennis) \cite{BB2, Tennis}, as well as for research scientists \cite{Scientists}.   A surprising observation is the
large numbers
of ``one-hit wonders", along with much smaller, but statistically significant and theoretically predictable, number of stellar ``iron-horse'' careers within these
competitive professions. We find a surprising statistical regularity which bridges the gap between the large number of individuals with very few career accomplishments and the few individuals with legendary career accomplishments. 
This empirical law emerges as a result of analyzing the entire population of agents/participants. 
Furthermore, by analyzing
the success rates 
across an entire population of agents, we quantify the time dependence of trends which alter the relative significance
of individual achievements. Analogous efforts are taking place in the bibliometric sciences in order  to establish
universal citation metrics which account for  variations across time and academic discipline 
\cite{Scientists,Growth8}. 

We demonstrate the utility of our detrending method by accounting for the changes in player performance over time in
professional baseball, which is particularly relevant to the induction process for HOF and to the debates regarding
the widespread use of PED in professional sports. There has also been debate over the use of cognitive enhancing drugs in academia \cite{CED1,CED2}. In baseball, we find a significant increase in home run rates in
recent years. Analyzing home run prowess,  we find a statistically significant 32\% increase in home run rates over the most recent 16-year ``steroid era'' period 
1994-2009 when compared to the previous 16-year period 1978-1993 (see Fig. \ref{prowesstrend}). Hence, the raw
accomplishments of sluggers during the steroids era will naturally supersede the records of sluggers from prior eras. So
how do we ensure that the legends of yesterday do not suffer from historical deflation? With the increased use of
sophisticated sabermetric statistics in  baseball and the recent application of network science methods to quantify the
extremely large number of head-to-head match-ups between pitcher and batter
\cite{BBnetworks}, a new picture of baseball is emerging \cite{BTWNums} which views all-time achievement in new light,
and is providing an objective framework for comparing achievements across time. 
In this paper, we consider the most natural measures for accomplishment, the statistics that are listed in every
box-score and on every
baseball card, in hope that the results are tangible to any historian or fan who is interested in 
reviewing and discussing the ``all-time greats''.

\acknowledgments{ We thank Sungho Han, Andrew West, and Peter Grassberger for their valuable comments.
 AMP and HES thank NSF for financial
support.  OP thanks NSERC and iCORE for financial support.  }

\newpage
\clearpage

\begin{widetext}

\begin{table}
\centering{{\normalsize
\begin{tabular}{|c|c||c||c|c|c|c||}
\hline
\textbf{\em Statistic x}  & $\langle r \rangle \pm \sigma$ &  $\langle x \rangle \pm  \sigma $  & $\alpha_{MLE}$ &
$x_{c}$  & $\sigma_{\Gamma}$ & $x^{*}$     \\
 \hline 
AB &  &$950 \pm 1800$ &  0.68 & 2900  & 1700   &   6500 (8000) \\
IPO & & $1300 \pm 2100$ & 0.57  & 3000  &  2000   &   7800 (9500) \\
K &  &$260 \pm 430 $ & 0.55  & 570  & 390   &   1500 (1800) \\
$K^{D}$ & $1.0 \pm 0.3$ &$240 \pm 400$ & 0.55  & 520 &  350  &   1400 (1700) \\
H  &  &$290 \pm 530$ &  0.65 & 820  & 490   &   1900 (2300) \\
$H^{D}$ & $1.0 \pm 0.04$ &$290 \pm 530$ & 0.65 & 820  & 490   &   1900 (2300) \\
W &  &$34 \pm 49$ & 0.40  & 55  & 43  &   170 (200) \\
$W^{D}$ & $1.0  \pm 0.01$ &$ 34 \pm 49$ &  0.40 & 55  & 43   &  170 (200) \\
HR  &  &$42 \pm 78$ & 0.53 & 89  & 62   &   240 (290) \\
$HR^{D}$ & $1.2 \pm 1.1$ &$36 \pm 65$ & 0.52  & 74  & 51   &   200 (240) \\
RBI & & $150 \pm 280$ &  0.61 & 390  & 240   &  940 (1200) \\
$RBI^{D}$ & $1.0 \pm 0.1$ &$150 \pm 270$ &  0.61 & 380  & 240   &   900 (1100) \\
 \hline
\end{tabular}
\caption{Parameter values for the pdfs of longevity and career metrics $X$ and detrended career metrics $X^{D}$ defined
in Eq. \ref{xdtotal}. The values of $\alpha$ and $x_{c} = \langle x
\rangle / (1-\alpha)$  are calculated using the 
gamma distribution maximum likelihood estimator. The values of the average and standard deviation $\langle x \rangle \pm
\sigma$ corresponds to actual data, while $\sigma_{\Gamma} = x_{c} \sqrt{1-\alpha}$ is the standard deviation of the 
gamma distribution ${\rm Gamma}(x; \alpha, x_{c})$. The benchmark value $x^{*}$ can be approximated as $x^{*} \approx 4 \
\sigma_{\Gamma}$ for the $f=0.02$ significance level \cite{BB2}. We also compute the value of $x^{*}$ for the $f=0.01$
significance level which we list in parenthesis. 
The first column corresponds to the average and standard deviation $\langle r \rangle \pm \sigma$ of
 the ratio $r \equiv X^{D}/X$ between an individual's detrended and traditional career success values.
}}}
\label{table:stats}
\end{table}

    \begin{center}
  \section*{Supplementary Tables}
  \end{center}
 
In Tables 2-11, we list top-50 tables for 5 career statistics and for 4 season statistics. For the two types of rankings, the columns are organized as follows:

\begin{itemize}
 \item[{\bf A}:] (Tables 2-6, Career statistics) The 4 columns on the left of each table list information for the ``traditional rank'' of career
statistics, where the top 50 players are ranked along with their final season (career length in seasons listed in
parenthesis) and their career metric tally. The 5  columns on the right of each table list information for the ``detrended rank'' ($Rank^{*}$) of career statistics, where the corresponding traditional rank (Rank) of the player is denoted in parenthesis.
 $L$ denotes the career length of the player.  The relative percent change  $\% Change = (Rank^{*}-Rank)/Rank$. 
  \item[{\bf B}:] (Tables 7-11, Season statistics) The 4 columns on the left  list the traditional ranking
of season statistics, where the top 50 players are ranked along with the year. The right columns list the detrended
ranking of season statistics $Rank^{*}$. $Y\#$ denotes the number of years into the career. The relative percent change  $\% Change = (Rank^{*}-Rank)/Rank$. 
\end{itemize}

\begin{table}
\centering{ {\footnotesize
\begin{tabular}{@{\vrule height .5 pt depth4pt  width0pt}lccc||lcccc}
&\multicolumn4c{{\bf Traditional Rank}}&\multicolumn4c{{\bf Detrended Rank}}\\
\noalign{
\vskip-1pt} Rank & Name & Final Season (L) & Career Metric & Rank$^{*}$(Rank)  &  \% Change & Name & Final Season (L)
& Career Metric \\
\hline 
1 & Barry  Bonds & 2007  (22)  &762 & 1(3)  & 66 & Babe  Ruth & 1935  (22)  &1215 \\
2 & Hank  Aaron & 1976  (23)  &755 & 2(23)  & 91 & Mel  Ott & 1947  (22)  &637 \\
3 & Babe  Ruth & 1935  (22)  &714 & 3(26)  & 88 & Lou  Gehrig & 1939  (17)  &635 \\
4 & Willie  Mays & 1973  (22)  &660 & 3(17)  & 82 & Jimmie  Foxx & 1945  (20)  &635 \\
5 & Ken  Griffey Jr. & 2009  (21)  &630 & 5(2)  & -150 & Hank  Aaron & 1976  (23)  &582 \\
6 & Sammy  Sosa & 2007  (18)  &609 & 6(124)  & 95 & Rogers  Hornsby & 1937  (23)  &528 \\
7 & Frank  Robinson & 1976  (21)  &586 & 7(192)  & 96 & Cy  Williams & 1930  (19)  &527 \\
8 & Alex  Rodriguez & 2009  (16)  &583 & 8(1)  & -700 & Barry  Bonds & 2007  (22)  &502 \\
8 & Mark  McGwire & 2001  (16)  &583 & 9(4)  & -125 & Willie  Mays & 1973  (22)  &490 \\
10 & Harmon  Killebrew & 1975  (22)  &573 & 10(18)  & 44 & Ted  Williams & 1960  (19)  &482 \\
11 & Rafael  Palmeiro & 2005  (20)  &569 & 11(13)  & 15 & Reggie  Jackson & 1987  (21)  &478 \\
12 & Jim  Thome & 2009  (19)  &564 & 12(14)  & 14 & Mike  Schmidt & 1989  (18)  &463 \\
13 & Reggie  Jackson & 1987  (21)  &563 & 13(7)  & -85 & Frank  Robinson & 1976  (21)  &444 \\
14 & Mike  Schmidt & 1989  (18)  &548 & 14(10)  & -40 & Harmon  Killebrew & 1975  (22)  &437 \\
15 & Manny  Ramirez & 2009  (17)  &546 & 15(577)  & 97 & Gavvy  Cravath & 1920  (11)  &433 \\
16 & Mickey  Mantle & 1968  (18)  &536 & 16(718)  & 97 & Honus  Wagner & 1917  (21)  &420 \\
17 & Jimmie  Foxx & 1945  (20)  &534 & 17(18)  & 5 & Willie  McCovey & 1980  (22)  &417 \\
18 & Ted  Williams & 1960  (19)  &521 & 18(557)  & 96 & Harry  Stovey & 1893  (14)  &413 \\
18 & Frank  Thomas & 2008  (19)  &521 & 19(5)  & -280 & Ken  Griffey Jr. & 2009  (21)  &411 \\
18 & Willie  McCovey & 1980  (22)  &521 & 20(28)  & 28 & Stan  Musial & 1963  (22)  &410 \\
21 & Eddie  Mathews & 1968  (17)  &512 & 21(28)  & 25 & Willie  Stargell & 1982  (21)  &399 \\
21 & Ernie  Banks & 1971  (19)  &512 & 22(25)  & 12 & Eddie  Murray & 1997  (21)  &397 \\
23 & Mel  Ott & 1947  (22)  &511 & 22(8)  & -175 & Mark  McGwire & 2001  (16)  &397 \\
24 & Gary  Sheffield & 2009  (22)  &509 & 24(16)  & -50 & Mickey  Mantle & 1968  (18)  &394 \\
25 & Eddie  Murray & 1997  (21)  &504 & 24(113)  & 78 & Al  Simmons & 1944  (20)  &394 \\
26 & Fred  McGriff & 2004  (19)  &493 & 26(470)  & 94 & Roger  Connor & 1897  (18)  &389 \\
26 & Lou  Gehrig & 1939  (17)  &493 & 27(752)  & 96 & Sam  Crawford & 1917  (19)  &386 \\
28 & Willie  Stargell & 1982  (21)  &475 & 28(70)  & 60 & Joe  DiMaggio & 1951  (13)  &382 \\
28 & Stan  Musial & 1963  (22)  &475 & 29(6)  & -383 & Sammy  Sosa & 2007  (18)  &381 \\
30 & Carlos  Delgado & 2009  (17)  &473 & 30(35)  & 14 & Dave  Kingman & 1986  (16)  &380 \\
31 & Dave  Winfield & 1995  (22)  &465 & 30(31)  & 3 & Dave  Winfield & 1995  (22)  &380 \\
32 & Jose  Canseco & 2001  (17)  &462 & 32(21)  & -52 & Ernie  Banks & 1971  (19)  &377 \\
33 & Carl  Yastrzemski & 1983  (23)  &452 & 33(91)  & 63 & Hank  Greenberg & 1947  (13)  &376 \\
34 & Jeff  Bagwell & 2005  (15)  &449 & 34(72)  & 52 & Johnny  Mize & 1953  (15)  &375 \\
35 & Dave  Kingman & 1986  (16)  &442 & 34(21)  & -61 & Eddie  Mathews & 1968  (17)  &375 \\
36 & Andre  Dawson & 1996  (21)  &438 & 36(33)  & -9 & Carl  Yastrzemski & 1983  (23)  &373 \\
37 & Juan  Gonzalez & 2005  (17)  &434 & 36(594)  & 93 & Ty  Cobb & 1928  (24)  &373 \\
38 & Cal  Ripken & 2001  (21)  &431 & 38(11)  & -245 & Rafael  Palmeiro & 2005  (20)  &368 \\
39 & Mike  Piazza & 2007  (16)  &427 & 39(125)  & 68 & Chuck  Klein & 1944  (17)  &361 \\
40 & Billy  Williams & 1976  (18)  &426 & 40(999)  & 95 & Harry  Davis & 1917  (22)  &360 \\
40 & Chipper  Jones & 2009  (16)  &426 & 41(667)  & 93 & Dan  Brouthers & 1904  (19)  &356 \\
42 & Darrell  Evans & 1989  (21)  &414 & 42(8)  & -425 & Alex  Rodriguez & 2009  (16)  &355 \\
43 & Jason  Giambi & 2009  (15)  &409 & 43(802)  & 94 & Frank  Schulte & 1918  (15)  &351 \\
44 & Duke  Snider & 1964  (18)  &407 & 44(135)  & 67 & Bob  Johnson & 1945  (13)  &348 \\
44 & Vladimir  Guerrero & 2009  (14)  &407 & 44(36)  & -22 & Andre  Dawson & 1996  (21)  &348 \\
46 & Al  Kaline & 1974  (22)  &399 & 46(12)  & -283 & Jim  Thome & 2009  (19)  &345 \\
46 & Andres  Galarraga & 2004  (19)  &399 & 46(26)  & -76 & Fred  McGriff & 2004  (19)  &345 \\
48 & Dale  Murphy & 1993  (18)  &398 & 46(40)  & -15 & Billy  Williams & 1976  (18)  &345 \\
49 & Joe  Carter & 1998  (16)  &396 & 49(153)  & 67 & Rudy  York & 1948  (13)  &343 \\
50 & Graig  Nettles & 1988  (22)  &390 & 50(42)  & -19 & Darrell  Evans & 1989  (21)  &342 \\
\hline
\end{tabular}}}
\caption{  Ranking of Career Home Runs (1871 - 2009). 
 }
\label{table:careerHR}
\end{table}

\begin{table}[h]
\centering{ {\footnotesize
\begin{tabular}{@{\vrule height .5 pt depth4pt  width0pt}lccc||lcccc}
&\multicolumn4c{{\bf Traditional Rank}}&\multicolumn4c{{\bf Detrended Rank}}\\
\noalign{
\vskip-1pt} Rank & Name & Final Season (L) & Career Metric & Rank$^{*}$(Rank)  &  \% Change & Name & Final Season (L)
& Career Metric \\
\hline 
1 & Pete  Rose & 1986  (24)  &4256 & 1(1)  & 0 & Pete  Rose & 1986  (24)  &4409 \\
2 & Ty  Cobb & 1928  (24)  &4189 & 2(2)  & 0 & Ty  Cobb & 1928  (24)  &4166 \\
3 & Hank  Aaron & 1976  (23)  &3771 & 3(3)  & 0 & Hank  Aaron & 1976  (23)  &3890 \\
4 & Stan  Musial & 1963  (22)  &3630 & 4(4)  & 0 & Stan  Musial & 1963  (22)  &3661 \\
5 & Tris  Speaker & 1928  (22)  &3514 & 5(6)  & 16 & Carl  Yastrzemski & 1983  (23)  &3537 \\
6 & Carl  Yastrzemski & 1983  (23)  &3419 & 6(8)  & 25 & Honus  Wagner & 1917  (21)  &3484 \\
7 & Cap  Anson & 1897  (27)  &3418 & 7(7)  & 0 & Cap  Anson & 1897  (27)  &3464 \\
8 & Honus  Wagner & 1917  (21)  &3415 & 8(5)  & -60 & Tris  Speaker & 1928  (22)  &3449 \\
9 & Paul  Molitor & 1998  (21)  &3319 & 9(11)  & 18 & Willie  Mays & 1973  (22)  &3375 \\
10 & Eddie  Collins & 1930  (25)  &3315 & 10(9)  & -11 & Paul  Molitor & 1998  (21)  &3361 \\
11 & Willie  Mays & 1973  (22)  &3283 & 11(12)  & 8 & Eddie  Murray & 1997  (21)  &3303 \\
12 & Eddie  Murray & 1997  (21)  &3255 & 12(13)  & 7 & Nap  Lajoie & 1916  (21)  &3291 \\
13 & Nap  Lajoie & 1916  (21)  &3242 & 13(10)  & -30 & Eddie  Collins & 1930  (25)  &3266 \\
14 & Cal  Ripken & 2001  (21)  &3184 & 14(15)  & 6 & George  Brett & 1993  (21)  &3222 \\
15 & George  Brett & 1993  (21)  &3154 & 15(14)  & -7 & Cal  Ripken & 2001  (21)  &3219 \\
16 & Paul  Waner & 1945  (20)  &3152 & 16(17)  & 5 & Robin  Yount & 1993  (20)  &3209 \\
17 & Robin  Yount & 1993  (20)  &3142 & 17(18)  & 5 & Tony  Gwynn & 2001  (20)  &3175 \\
18 & Tony  Gwynn & 2001  (20)  &3141 & 18(19)  & 5 & Dave  Winfield & 1995  (22)  &3171 \\
19 & Dave  Winfield & 1995  (22)  &3110 & 19(23)  & 17 & Lou  Brock & 1979  (19)  &3150 \\
20 & Craig  Biggio & 2007  (20)  &3060 & 20(22)  & 9 & Rod  Carew & 1985  (19)  &3149 \\
21 & Rickey  Henderson & 2003  (25)  &3055 & 21(27)  & 22 & Roberto  Clemente & 1972  (18)  &3107 \\
22 & Rod  Carew & 1985  (19)  &3053 & 22(26)  & 15 & Al  Kaline & 1974  (22)  &3094 \\
23 & Lou  Brock & 1979  (19)  &3023 & 23(21)  & -9 & Rickey  Henderson & 2003  (25)  &3089 \\
24 & Rafael  Palmeiro & 2005  (20)  &3020 & 24(20)  & -20 & Craig  Biggio & 2007  (20)  &3060 \\
25 & Wade  Boggs & 1999  (18)  &3010 & 25(25)  & 0 & Wade  Boggs & 1999  (18)  &3053 \\
26 & Al  Kaline & 1974  (22)  &3007 & 26(29)  & 10 & Sam  Crawford & 1917  (19)  &3046 \\
27 & Roberto  Clemente & 1972  (18)  &3000 & 27(30)  & 10 & Frank  Robinson & 1976  (21)  &3040 \\
28 & Sam  Rice & 1934  (20)  &2987 & 28(24)  & -16 & Rafael  Palmeiro & 2005  (20)  &3034 \\
29 & Sam  Crawford & 1917  (19)  &2961 & 29(16)  & -81 & Paul  Waner & 1945  (20)  &2968 \\
30 & Frank  Robinson & 1976  (21)  &2943 & 30(42)  & 28 & Brooks  Robinson & 1977  (23)  &2955 \\
31 & Barry  Bonds & 2007  (22)  &2935 & 31(31)  & 0 & Barry  Bonds & 2007  (22)  &2948 \\
32 & Willie  Keeler & 1910  (19)  &2932 & 32(33)  & 3 & Jake  Beckley & 1907  (20)  &2905 \\
33 & Rogers  Hornsby & 1937  (23)  &2930 & 33(40)  & 17 & Harold  Baines & 2001  (22)  &2900 \\
33 & Jake  Beckley & 1907  (20)  &2930 & 34(32)  & -6 & Willie  Keeler & 1910  (19)  &2872 \\
35 & Al  Simmons & 1944  (20)  &2927 & 35(47)  & 25 & Vada  Pinson & 1975  (18)  &2863 \\
36 & Zack  Wheat & 1927  (19)  &2884 & 36(52)  & 30 & Tony  Perez & 1986  (23)  &2831 \\
37 & Frankie  Frisch & 1937  (19)  &2880 & 37(58)  & 36 & Billy  Williams & 1976  (18)  &2830 \\
38 & Mel  Ott & 1947  (22)  &2876 & 38(45)  & 15 & Andre  Dawson & 1996  (21)  &2823 \\
39 & Babe  Ruth & 1935  (22)  &2873 & 39(55)  & 29 & Rusty  Staub & 1985  (23)  &2821 \\
40 & Harold  Baines & 2001  (22)  &2866 & 40(50)  & 20 & Al  Oliver & 1985  (18)  &2813 \\
41 & Jesse  Burkett & 1905  (16)  &2850 & 41(36)  & -13 & Zack  Wheat & 1927  (19)  &2809 \\
42 & Brooks  Robinson & 1977  (23)  &2848 & 42(28)  & -50 & Sam  Rice & 1934  (20)  &2794 \\
43 & Charlie  Gehringer & 1942  (19)  &2839 & 43(56)  & 23 & Bill  Buckner & 1990  (22)  &2779 \\
44 & George  Sisler & 1930  (15)  &2812 & 44(63)  & 30 & Luis  Aparicio & 1973  (18)  &2771 \\
45 & Andre  Dawson & 1996  (21)  &2774 & 45(57)  & 21 & Dave  Parker & 1991  (19)  &2770 \\
46 & Ken  Griffey Jr. & 2009  (21)  &2763 & 46(41)  & -12 & Jesse  Burkett & 1905  (16)  &2768 \\
47 & Vada  Pinson & 1975  (18)  &2757 & 47(33)  & -42 & Rogers  Hornsby & 1937  (23)  &2766 \\
48 & Luke  Appling & 1950  (20)  &2749 & 47(46)  & -2 & Ken  Griffey Jr. & 2009  (21)  &2766 \\
49 & Derek  Jeter & 2009  (15)  &2747 & 49(38)  & -28 & Mel  Ott & 1947  (22)  &2745 \\
50 & Al  Oliver & 1985  (18)  &2743 & 50(53)  & 5 & Roberto  Alomar & 2004  (17)  &2738 \\
\hline
\end{tabular}}}
\caption{ Ranking of Career Hits (1871 - 2009). }
\label{table:careerH}
\end{table}

\begin{table}[h]
\centering{ {\footnotesize
\begin{tabular}{@{\vrule height .5 pt depth4pt  width0pt}lccc||lcccc}
&\multicolumn4c{{\bf Traditional Rank}}&\multicolumn4c{{\bf Detrended Rank}}\\
\noalign{
\vskip-1pt} Rank & Name & Final Season (L) & Career Metric & Rank$^{*}$(Rank)  &  \% Change & Name & Final Season (L)
& Career Metric \\
\hline 
1 & Hank  Aaron & 1976  (23)  &2297 & 1(1)  & 0 & Hank  Aaron & 1976  (23)  &2362 \\
2 & Babe  Ruth & 1935  (22)  &2217 & 2(3)  & 33 & Cap  Anson & 1897  (27)  &2295 \\
3 & Cap  Anson & 1897  (27)  &2076 & 3(7)  & 57 & Ty  Cobb & 1928  (24)  &2176 \\
4 & Barry  Bonds & 2007  (22)  &1996 & 4(2)  & -100 & Babe  Ruth & 1935  (22)  &2120 \\
5 & Lou  Gehrig & 1939  (17)  &1995 & 5(20)  & 75 & Honus  Wagner & 1917  (21)  &1986 \\
6 & Stan  Musial & 1963  (22)  &1951 & 6(12)  & 50 & Carl  Yastrzemski & 1983  (23)  &1933 \\
7 & Ty  Cobb & 1928  (24)  &1937 & 7(10)  & 30 & Willie  Mays & 1973  (22)  &1930 \\
8 & Jimmie  Foxx & 1945  (20)  &1922 & 8(6)  & -33 & Stan  Musial & 1963  (22)  &1911 \\
9 & Eddie  Murray & 1997  (21)  &1917 & 9(9)  & 0 & Eddie  Murray & 1997  (21)  &1899 \\
10 & Willie  Mays & 1973  (22)  &1903 & 10(18)  & 44 & Frank  Robinson & 1976  (21)  &1868 \\
11 & Mel  Ott & 1947  (22)  &1860 & 11(15)  & 26 & Dave  Winfield & 1995  (22)  &1848 \\
12 & Carl  Yastrzemski & 1983  (23)  &1844 & 11(4)  & -175 & Barry  Bonds & 2007  (22)  &1848 \\
13 & Ted  Williams & 1960  (19)  &1839 & 13(47)  & 72 & Sam  Crawford & 1917  (19)  &1798 \\
14 & Rafael  Palmeiro & 2005  (20)  &1835 & 14(5)  & -180 & Lou  Gehrig & 1939  (17)  &1796 \\
15 & Dave  Winfield & 1995  (22)  &1833 & 15(31)  & 51 & Nap  Lajoie & 1916  (21)  &1782 \\
16 & Ken  Griffey Jr. & 2009  (21)  &1829 & 16(13)  & -23 & Ted  Williams & 1960  (19)  &1776 \\
17 & Al  Simmons & 1944  (20)  &1827 & 17(23)  & 26 & Reggie  Jackson & 1987  (21)  &1770 \\
18 & Frank  Robinson & 1976  (21)  &1812 & 18(8)  & -125 & Jimmie  Foxx & 1945  (20)  &1747 \\
19 & Manny  Ramirez & 2009  (17)  &1788 & 19(27)  & 29 & Tony  Perez & 1986  (23)  &1745 \\
20 & Honus  Wagner & 1917  (21)  &1732 & 20(11)  & -81 & Mel  Ott & 1947  (22)  &1725 \\
21 & Alex  Rodriguez & 2009  (16)  &1706 & 21(14)  & -50 & Rafael  Palmeiro & 2005  (20)  &1689 \\
22 & Frank  Thomas & 2008  (19)  &1704 & 22(16)  & -37 & Ken  Griffey Jr. & 2009  (21)  &1679 \\
23 & Reggie  Jackson & 1987  (21)  &1702 & 23(28)  & 17 & Ernie  Banks & 1971  (19)  &1661 \\
24 & Cal  Ripken & 2001  (21)  &1695 & 24(45)  & 46 & Tris  Speaker & 1928  (22)  &1653 \\
25 & Gary  Sheffield & 2009  (22)  &1676 & 25(17)  & -47 & Al  Simmons & 1944  (20)  &1646 \\
26 & Sammy  Sosa & 2007  (18)  &1667 & 25(35)  & 28 & Harmon  Killebrew & 1975  (22)  &1646 \\
27 & Tony  Perez & 1986  (23)  &1652 & 27(42)  & 35 & Willie  Stargell & 1982  (21)  &1640 \\
28 & Ernie  Banks & 1971  (19)  &1636 & 28(40)  & 30 & Willie  McCovey & 1980  (22)  &1635 \\
29 & Harold  Baines & 2001  (22)  &1628 & 29(24)  & -20 & Cal  Ripken & 2001  (21)  &1625 \\
30 & Goose  Goslin & 1938  (18)  &1609 & 30(32)  & 6 & Mike  Schmidt & 1989  (18)  &1622 \\
31 & Nap  Lajoie & 1916  (21)  &1599 & 31(37)  & 16 & Al  Kaline & 1974  (22)  &1614 \\
32 & Mike  Schmidt & 1989  (18)  &1595 & 31(32)  & 3 & George  Brett & 1993  (21)  &1614 \\
32 & George  Brett & 1993  (21)  &1595 & 33(19)  & -73 & Manny  Ramirez & 2009  (17)  &1598 \\
34 & Andre  Dawson & 1996  (21)  &1591 & 34(34)  & 0 & Andre  Dawson & 1996  (21)  &1586 \\
35 & Harmon  Killebrew & 1975  (22)  &1584 & 35(29)  & -20 & Harold  Baines & 2001  (22)  &1573 \\
35 & Rogers  Hornsby & 1937  (23)  &1584 & 36(52)  & 30 & Billy  Williams & 1976  (18)  &1567 \\
37 & Al  Kaline & 1974  (22)  &1583 & 37(35)  & -5 & Rogers  Hornsby & 1937  (23)  &1557 \\
38 & Jake  Beckley & 1907  (20)  &1575 & 38(22)  & -72 & Frank  Thomas & 2008  (19)  &1554 \\
39 & Jim  Thome & 2009  (19)  &1565 & 39(53)  & 26 & Rusty  Staub & 1985  (23)  &1550 \\
40 & Willie  McCovey & 1980  (22)  &1555 & 40(25)  & -60 & Gary  Sheffield & 2009  (22)  &1533 \\
41 & Fred  McGriff & 2004  (19)  &1550 & 41(21)  & -95 & Alex  Rodriguez & 2009  (16)  &1528 \\
42 & Willie  Stargell & 1982  (21)  &1540 & 42(43)  & 2 & Harry  Heilmann & 1932  (17)  &1521 \\
43 & Harry  Heilmann & 1932  (17)  &1539 & 43(51)  & 15 & Dave  Parker & 1991  (19)  &1515 \\
44 & Joe  DiMaggio & 1951  (13)  &1537 & 44(38)  & -15 & Jake  Beckley & 1907  (20)  &1511 \\
45 & Tris  Speaker & 1928  (22)  &1529 & 45(26)  & -73 & Sammy  Sosa & 2007  (18)  &1504 \\
45 & Jeff  Bagwell & 2005  (15)  &1529 & 46(50)  & 8 & Mickey  Mantle & 1968  (18)  &1502 \\
47 & Sam  Crawford & 1917  (19)  &1525 & 47(56)  & 16 & Jim  Rice & 1989  (16)  &1473 \\
48 & Jeff  Kent & 2008  (17)  &1518 & 48(30)  & -60 & Goose  Goslin & 1938  (18)  &1456 \\
49 & Carlos  Delgado & 2009  (17)  &1512 & 49(103)  & 52 & Dan  Brouthers & 1904  (19)  &1449 \\
50 & Mickey  Mantle & 1968  (18)  &1509 & 50(73)  & 31 & Johnny  Bench & 1983  (17)  &1447 \\
\hline
\end{tabular}}}
\caption{ Ranking of Career Runs Batted In (1871 - 2009). }
\label{table:careerRBI}
\end{table}

\begin{table}[h]
\centering{ {\footnotesize
\begin{tabular}{@{\vrule height .5 pt depth4pt  width0pt}lccc||lcccc}
&\multicolumn4c{{\bf Traditional Rank}}&\multicolumn4c{{\bf Detrended Rank}}\\
\noalign{
\vskip-1pt} Rank & Name & Final Season (L) & Career Metric & Rank$^{*}$(Rank)  &  \% Change & Name & Final Season (L)
& Career Metric \\
\hline 
1 & Nolan  Ryan & 1993  (27)  &5714 & 1(1)  & 0 & Nolan  Ryan & 1993  (27)  &4937 \\
2 & Randy  Johnson & 2009  (22)  &4875 & 2(9)  & 77 & Walter  Johnson & 1927  (21)  &4681 \\
3 & Roger  Clemens & 2007  (24)  &4672 & 3(20)  & 85 & Cy  Young & 1911  (22)  &4216 \\
4 & Steve  Carlton & 1988  (24)  &4136 & 4(4)  & 0 & Steve  Carlton & 1988  (24)  &3615 \\
5 & Bert  Blyleven & 1992  (22)  &3701 & 5(2)  & -150 & Randy  Johnson & 2009  (22)  &3524 \\
6 & Tom  Seaver & 1986  (20)  &3640 & 6(3)  & -100 & Roger  Clemens & 2007  (24)  &3483 \\
7 & Don  Sutton & 1988  (23)  &3574 & 7(46)  & 84 & Lefty  Grove & 1941  (17)  &3307 \\
8 & Gaylord  Perry & 1983  (22)  &3534 & 8(27)  & 70 & Tim  Keefe & 1893  (14)  &3241 \\
9 & Walter  Johnson & 1927  (21)  &3509 & 9(5)  & -80 & Bert  Blyleven & 1992  (22)  &3223 \\
10 & Greg  Maddux & 2008  (23)  &3371 & 10(59)  & 83 & Dazzy  Vance & 1935  (16)  &3208 \\
11 & Phil  Niekro & 1987  (24)  &3342 & 11(26)  & 57 & Bob  Feller & 1956  (18)  &3193 \\
12 & Fergie  Jenkins & 1983  (19)  &3192 & 12(6)  & -100 & Tom  Seaver & 1986  (20)  &3154 \\
13 & Pedro  Martinez & 2009  (18)  &3154 & 13(75)  & 82 & Amos  Rusie & 1901  (10)  &3138 \\
14 & Bob  Gibson & 1975  (17)  &3117 & 14(29)  & 51 & Christy  Mathewson & 1916  (17)  &3116 \\
15 & Curt  Schilling & 2007  (20)  &3116 & 15(7)  & -114 & Don  Sutton & 1988  (23)  &3073 \\
16 & John  Smoltz & 2009  (21)  &3084 & 16(43)  & 62 & Rube  Waddell & 1910  (13)  &3054 \\
17 & Jim  Bunning & 1971  (17)  &2855 & 17(79)  & 78 & Kid  Nichols & 1906  (15)  &3042 \\
18 & Mickey  Lolich & 1979  (16)  &2832 & 18(8)  & -125 & Gaylord  Perry & 1983  (22)  &3015 \\
19 & Mike  Mussina & 2008  (18)  &2813 & 19(52)  & 63 & Pete  Alexander & 1930  (20)  &2926 \\
20 & Cy  Young & 1911  (22)  &2803 & 20(11)  & -81 & Phil  Niekro & 1987  (24)  &2916 \\
21 & Frank  Tanana & 1993  (21)  &2773 & 21(223)  & 90 & Bobby  Mathews & 1887  (15)  &2871 \\
22 & David  Cone & 2003  (17)  &2668 & 22(66)  & 66 & Red  Ruffing & 1947  (22)  &2804 \\
23 & Chuck  Finley & 2002  (17)  &2610 & 23(48)  & 52 & Eddie  Plank & 1917  (17)  &2748 \\
24 & Tom  Glavine & 2008  (22)  &2607 & 24(12)  & -100 & Fergie  Jenkins & 1983  (19)  &2740 \\
25 & Warren  Spahn & 1965  (21)  &2583 & 25(57)  & 56 & Bobo  Newsom & 1953  (20)  &2714 \\
26 & Bob  Feller & 1956  (18)  &2581 & 26(25)  & -4 & Warren  Spahn & 1965  (21)  &2660 \\
27 & Tim  Keefe & 1893  (14)  &2562 & 27(14)  & -92 & Bob  Gibson & 1975  (17)  &2534 \\
28 & Jerry  Koosman & 1985  (19)  &2556 & 28(119)  & 76 & Gus  Weyhing & 1901  (14)  &2533 \\
29 & Christy  Mathewson & 1916  (17)  &2502 & 29(68)  & 57 & John  Clarkson & 1894  (12)  &2522 \\
30 & Don  Drysdale & 1969  (14)  &2486 & 30(10)  & -200 & Greg  Maddux & 2008  (23)  &2475 \\
31 & Jack  Morris & 1994  (18)  &2478 & 31(42)  & 26 & Early  Wynn & 1963  (23)  &2471 \\
32 & Mark  Langston & 1999  (16)  &2464 & 32(21)  & -52 & Frank  Tanana & 1993  (21)  &2417 \\
33 & Jim  Kaat & 1983  (25)  &2461 & 32(84)  & 61 & Pud  Galvin & 1892  (15)  &2417 \\
34 & Sam  McDowell & 1975  (15)  &2453 & 34(17)  & -100 & Jim  Bunning & 1971  (17)  &2373 \\
35 & Luis  Tiant & 1982  (19)  &2416 & 35(163)  & 78 & Burleigh  Grimes & 1934  (19)  &2361 \\
36 & Dennis  Eckersley & 1998  (24)  &2401 & 36(40)  & 10 & Robin  Roberts & 1966  (19)  &2355 \\
37 & Kevin  Brown & 2005  (19)  &2397 & 37(122)  & 69 & Vic  Willis & 1910  (13)  &2352 \\
38 & Sandy  Koufax & 1966  (12)  &2396 & 38(18)  & -111 & Mickey  Lolich & 1979  (16)  &2334 \\
39 & Charlie  Hough & 1994  (25)  &2362 & 39(85)  & 54 & Tony  Mullane & 1894  (13)  &2320 \\
40 & Robin  Roberts & 1966  (19)  &2357 & 40(113)  & 64 & Carl  Hubbell & 1943  (16)  &2312 \\
41 & Jamie  Moyer & 2009  (23)  &2342 & 41(106)  & 61 & Jim  McCormick & 1887  (10)  &2301 \\
42 & Early  Wynn & 1963  (23)  &2334 & 42(86)  & 51 & Hal  Newhouser & 1955  (17)  &2275 \\
43 & Rube  Waddell & 1910  (13)  &2316 & 43(129)  & 66 & Jack  Powell & 1912  (16)  &2266 \\
44 & Juan  Marichal & 1975  (16)  &2303 & 44(80)  & 45 & Mickey  Welch & 1892  (13)  &2263 \\
45 & Dwight  Gooden & 2000  (16)  &2293 & 45(28)  & -60 & Jerry  Koosman & 1985  (19)  &2258 \\
46 & Lefty  Grove & 1941  (17)  &2266 & 46(16)  & -187 & John  Smoltz & 2009  (21)  &2256 \\
47 & Javier  Vazquez & 2009  (12)  &2253 & 47(114)  & 58 & Tommy  Bridges & 1946  (16)  &2247 \\
48 & Eddie  Plank & 1917  (17)  &2246 & 47(82)  & 42 & Charley  Radbourn & 1891  (11)  &2247 \\
49 & Tommy  John & 1989  (26)  &2245 & 49(13)  & -276 & Pedro  Martinez & 2009  (18)  &2225 \\
50 & Jim  Palmer & 1984  (19)  &2212 & 50(15)  & -233 & Curt  Schilling & 2007  (20)  &2222 \\
\hline
\end{tabular}}}
\caption{ Ranking of Career Strikeouts (1871 - 2009).  }
\label{table:careerK}
\end{table}

\begin{table}[h]
\centering{ {\footnotesize
\begin{tabular}{@{\vrule height .5 pt depth4pt  width0pt}lccc||lcccc}
&\multicolumn4c{{\bf Traditional Rank}}&\multicolumn4c{{\bf Detrended Rank}}\\
\noalign{
\vskip-1pt} Rank & Name & Final Season (L) & Career Metric & Rank$^{*}$(Rank)  &  \% Change & Name & Final Season (L)
& Career Metric \\
\hline 
1 & Cy  Young & 1911  (22)  &511 & 1(1)  & 0 & Cy  Young & 1911  (22)  &510 \\
2 & Walter  Johnson & 1927  (21)  &417 & 2(2)  & 0 & Walter  Johnson & 1927  (21)  &420 \\
3 & Christy  Mathewson & 1916  (17)  &373 & 3(3)  & 0 & Christy  Mathewson & 1916  (17)  &376 \\
3 & Pete  Alexander & 1930  (20)  &373 & 4(3)  & -33 & Pete  Alexander & 1930  (20)  &375 \\
5 & Pud  Galvin & 1892  (15)  &364 & 5(5)  & 0 & Pud  Galvin & 1892  (15)  &365 \\
6 & Warren  Spahn & 1965  (21)  &363 & 6(6)  & 0 & Warren  Spahn & 1965  (21)  &362 \\
7 & Kid  Nichols & 1906  (15)  &361 & 7(7)  & 0 & Kid  Nichols & 1906  (15)  &359 \\
8 & Greg  Maddux & 2008  (23)  &355 & 8(8)  & 0 & Greg  Maddux & 2008  (23)  &351 \\
9 & Roger  Clemens & 2007  (24)  &354 & 9(9)  & 0 & Roger  Clemens & 2007  (24)  &350 \\
10 & Tim  Keefe & 1893  (14)  &342 & 10(10)  & 0 & Tim  Keefe & 1893  (14)  &342 \\
11 & Steve  Carlton & 1988  (24)  &329 & 11(11)  & 0 & Steve  Carlton & 1988  (24)  &329 \\
12 & John  Clarkson & 1894  (12)  &328 & 12(13)  & 7 & Eddie  Plank & 1917  (17)  &328 \\
13 & Eddie  Plank & 1917  (17)  &326 & 13(12)  & -8 & John  Clarkson & 1894  (12)  &327 \\
14 & Don  Sutton & 1988  (23)  &324 & 14(14)  & 0 & Don  Sutton & 1988  (23)  &324 \\
14 & Nolan  Ryan & 1993  (27)  &324 & 14(14)  & 0 & Nolan  Ryan & 1993  (27)  &324 \\
16 & Phil  Niekro & 1987  (24)  &318 & 16(16)  & 0 & Phil  Niekro & 1987  (24)  &318 \\
17 & Gaylord  Perry & 1983  (22)  &314 & 17(17)  & 0 & Gaylord  Perry & 1983  (22)  &314 \\
18 & Tom  Seaver & 1986  (20)  &311 & 18(18)  & 0 & Tom  Seaver & 1986  (20)  &311 \\
19 & Charley  Radbourn & 1891  (11)  &309 & 19(19)  & 0 & Charley  Radbourn & 1891  (11)  &308 \\
20 & Mickey  Welch & 1892  (13)  &307 & 20(20)  & 0 & Mickey  Welch & 1892  (13)  &307 \\
21 & Tom  Glavine & 2008  (22)  &305 & 21(21)  & 0 & Tom  Glavine & 2008  (22)  &302 \\
22 & Randy  Johnson & 2009  (22)  &303 & 22(25)  & 12 & Bobby  Mathews & 1887  (15)  &300 \\
23 & Early  Wynn & 1963  (23)  &300 & 22(23)  & 4 & Early  Wynn & 1963  (23)  &300 \\
23 & Lefty  Grove & 1941  (17)  &300 & 24(23)  & -4 & Lefty  Grove & 1941  (17)  &299 \\
25 & Bobby  Mathews & 1887  (15)  &297 & 24(22)  & -9 & Randy  Johnson & 2009  (22)  &299 \\
26 & Tommy  John & 1989  (26)  &288 & 26(26)  & 0 & Tommy  John & 1989  (26)  &288 \\
27 & Bert  Blyleven & 1992  (22)  &287 & 27(27)  & 0 & Bert  Blyleven & 1992  (22)  &287 \\
28 & Robin  Roberts & 1966  (19)  &286 & 28(28)  & 0 & Robin  Roberts & 1966  (19)  &285 \\
29 & Tony  Mullane & 1894  (13)  &284 & 29(29)  & 0 & Fergie  Jenkins & 1983  (19)  &284 \\
29 & Fergie  Jenkins & 1983  (19)  &284 & 30(31)  & 3 & Jim  Kaat & 1983  (25)  &283 \\
31 & Jim  Kaat & 1983  (25)  &283 & 30(29)  & -3 & Tony  Mullane & 1894  (13)  &283 \\
32 & Red  Ruffing & 1947  (22)  &273 & 32(32)  & 0 & Red  Ruffing & 1947  (22)  &272 \\
33 & Mike  Mussina & 2008  (18)  &270 & 33(33)  & 0 & Burleigh  Grimes & 1934  (19)  &270 \\
33 & Burleigh  Grimes & 1934  (19)  &270 & 34(35)  & 2 & Jim  Palmer & 1984  (19)  &268 \\
35 & Jim  Palmer & 1984  (19)  &268 & 35(33)  & -6 & Mike  Mussina & 2008  (18)  &267 \\
36 & Eppa  Rixey & 1933  (21)  &266 & 36(36)  & 0 & Eppa  Rixey & 1933  (21)  &266 \\
36 & Bob  Feller & 1956  (18)  &266 & 36(38)  & 5 & Jim  McCormick & 1887  (10)  &266 \\
38 & Jim  McCormick & 1887  (10)  &265 & 38(36)  & -5 & Bob  Feller & 1956  (18)  &265 \\
39 & Gus  Weyhing & 1901  (14)  &264 & 39(39)  & 0 & Gus  Weyhing & 1901  (14)  &263 \\
40 & Ted  Lyons & 1946  (21)  &260 & 40(40)  & 0 & Ted  Lyons & 1946  (21)  &259 \\
41 & Jamie  Moyer & 2009  (23)  &258 & 40(44)  & 9 & Al  Spalding & 1877  (7)  &259 \\
42 & Jack  Morris & 1994  (18)  &254 & 42(42)  & 0 & Red  Faber & 1933  (20)  &255 \\
42 & Red  Faber & 1933  (20)  &254 & 43(41)  & -4 & Jamie  Moyer & 2009  (23)  &254 \\
44 & Al  Spalding & 1877  (7)  &253 & 44(42)  & -4 & Jack  Morris & 1994  (18)  &253 \\
44 & Carl  Hubbell & 1943  (16)  &253 & 44(44)  & 0 & Carl  Hubbell & 1943  (16)  &253 \\
46 & Bob  Gibson & 1975  (17)  &251 & 46(46)  & 0 & Bob  Gibson & 1975  (17)  &251 \\
47 & Vic  Willis & 1910  (13)  &249 & 47(47)  & 0 & Vic  Willis & 1910  (13)  &250 \\
48 & Jack  Quinn & 1933  (23)  &247 & 48(48)  & 0 & Jack  Quinn & 1933  (23)  &247 \\
49 & Joe  McGinnity & 1908  (10)  &246 & 48(49)  & 2 & Joe  McGinnity & 1908  (10)  &247 \\
50 & Amos  Rusie & 1901  (10)  &245 & 50(50)  & 0 & Jack  Powell & 1912  (16)  &245 \\
\hline
\end{tabular}}}
\caption{ Ranking of Career Wins  (1890 - 2009). }
\label{table:careerW}
\end{table}

\begin{table}[h]
\centering{ {\footnotesize
\begin{tabular}{@{\vrule height .5 pt depth4pt  width0pt}lccc||lcccc}
&\multicolumn4c{{\bf Traditional Rank}}&\multicolumn4c{{\bf Detrended Rank}}\\
\noalign{
\vskip-1pt} Rank & Name & Season ($Y\#$) & Season Metric & Rank$^{*}$(Rank)  &  \% Change & Name & Season ($Y\#$) &
Season Metric \\
\hline 
1 & Barry  Bonds & 2001  (16)  &73 & 1(19)  & 94 & Babe  Ruth & 1920  (7)  &133 \\
2 & Mark  McGwire & 1998  (13)  &70 & 2(8)  & 75 & Babe  Ruth & 1927  (14)  &102 \\
3 & Sammy  Sosa & 1998  (10)  &66 & 3(9)  & 66 & Babe  Ruth & 1921  (8)  &100 \\
4 & Mark  McGwire & 1999  (14)  &65 & 4(72)  & 94 & Babe  Ruth & 1926  (13)  &82 \\
5 & Sammy  Sosa & 2001  (13)  &64 & 5(94)  & 94 & Babe  Ruth & 1924  (11)  &80 \\
6 & Sammy  Sosa & 1999  (11)  &63 & 5(72)  & 93 & Lou  Gehrig & 1927  (5)  &80 \\
7 & Roger  Maris & 1961  (5)  &61 & 7(19)  & 63 & Babe  Ruth & 1928  (15)  &77 \\
8 & Babe  Ruth & 1927  (14)  &60 & 8(61)  & 86 & Jimmie  Foxx & 1933  (9)  &70 \\
9 & Babe  Ruth & 1921  (8)  &59 & 9(94)  & 90 & Babe  Ruth & 1931  (18)  &68 \\
10 & Mark  McGwire & 1997  (12)  &58 & 9(94)  & 90 & Lou  Gehrig & 1931  (9)  &68 \\
10 & Ryan  Howard & 2006  (3)  &58 & 11(10)  & -10 & Jimmie  Foxx & 1932  (8)  &67 \\
10 & Hank  Greenberg & 1938  (7)  &58 & 12(215)  & 94 & Cy  Williams & 1923  (12)  &66 \\
10 & Jimmie  Foxx & 1932  (8)  &58 & 12(215)  & 94 & Babe  Ruth & 1923  (10)  &66 \\
14 & Alex  Rodriguez & 2002  (9)  &57 & 14(181)  & 92 & Rogers  Hornsby & 1922  (8)  &62 \\
14 & Luis  Gonzalez & 2001  (12)  &57 & 15(10)  & -50 & Hank  Greenberg & 1938  (7)  &60 \\
16 & Hack  Wilson & 1930  (8)  &56 & 16(301)  & 94 & Ken  Williams & 1922  (7)  &58 \\
16 & Ken  Griffey Jr. & 1998  (10)  &56 & 16(592)  & 97 & Rudy  York & 1943  (8)  &58 \\
16 & Ken  Griffey Jr. & 1997  (9)  &56 & 18(42)  & 57 & Lou  Gehrig & 1936  (14)  &57 \\
19 & Babe  Ruth & 1928  (15)  &54 & 18(42)  & 57 & Lou  Gehrig & 1934  (12)  &57 \\
19 & Babe  Ruth & 1920  (7)  &54 & 20(16)  & -25 & Hack  Wilson & 1930  (8)  &56 \\
19 & Alex  Rodriguez & 2007  (14)  &54 & 21(135)  & 84 & Hank  Greenberg & 1946  (12)  &55 \\
19 & David  Ortiz & 2006  (10)  &54 & 21(401)  & 94 & Tilly  Walker & 1922  (12)  &55 \\
19 & Mickey  Mantle & 1961  (11)  &54 & 23(94)  & 75 & Babe  Ruth & 1929  (16)  &53 \\
19 & Ralph  Kiner & 1949  (4)  &54 & 23(899)  & 97 & Charlie  Keller & 1943  (5)  &53 \\
25 & Jim  Thome & 2002  (12)  &52 & 25(301)  & 91 & Rogers  Hornsby & 1925  (11)  &52 \\
25 & Alex  Rodriguez & 2001  (8)  &52 & 25(36)  & 30 & Jimmie  Foxx & 1938  (14)  &52 \\
25 & Mark  McGwire & 1996  (11)  &52 & 25(519)  & 95 & Babe  Ruth & 1922  (9)  &52 \\
25 & Willie  Mays & 1965  (14)  &52 & 28(135)  & 79 & Jimmie  Foxx & 1934  (10)  &51 \\
25 & Mickey  Mantle & 1956  (6)  &52 & 28(1023)  & 97 & Hack  Wilson & 1927  (5)  &51 \\
25 & George  Foster & 1977  (9)  &52 & 28(1023)  & 97 & Cy  Williams & 1927  (16)  &51 \\
31 & Johnny  Mize & 1947  (9)  &51 & 28(457)  & 93 & Ted  Williams & 1942  (4)  &51 \\
31 & Willie  Mays & 1955  (4)  &51 & 32(161)  & 80 & Chuck  Klein & 1929  (2)  &50 \\
31 & Ralph  Kiner & 1947  (2)  &51 & 32(31)  & -3 & Johnny  Mize & 1947  (9)  &50 \\
31 & Andruw  Jones & 2005  (10)  &51 & 32(31)  & -3 & Ralph  Kiner & 1947  (2)  &50 \\
31 & Cecil  Fielder & 1990  (5)  &51 & 35(94)  & 62 & Joe  DiMaggio & 1937  (2)  &49 \\
36 & Greg  Vaughn & 1998  (10)  &50 & 35(592)  & 94 & Babe  Ruth & 1933  (20)  &49 \\
36 & Sammy  Sosa & 2000  (12)  &50 & 35(42)  & 16 & Babe  Ruth & 1930  (17)  &49 \\
36 & Jimmie  Foxx & 1938  (14)  &50 & 35(1134)  & 96 & Bill  Nicholson & 1943  (6)  &49 \\
36 & Prince  Fielder & 2007  (3)  &50 & 35(181)  & 80 & Hal  Trosky & 1936  (4)  &49 \\
36 & Albert  Belle & 1995  (7)  &50 & 35(686)  & 94 & Bill  Nicholson & 1944  (7)  &49 \\
36 & Brady  Anderson & 1996  (9)  &50 & 41(181)  & 77 & Mel  Ott & 1929  (4)  &48 \\
42 & Larry  Walker & 1997  (9)  &49 & 41(19)  & -115 & Ralph  Kiner & 1949  (4)  &48 \\
42 & Jim  Thome & 2001  (11)  &49 & 41(215)  & 80 & Jimmie  Foxx & 1936  (12)  &48 \\
42 & Sammy  Sosa & 2002  (14)  &49 & 41(353)  & 88 & Ted  Williams & 1946  (5)  &48 \\
42 & Babe  Ruth & 1930  (17)  &49 & 45(215)  & 79 & Babe  Ruth & 1932  (19)  &47 \\
42 & Frank  Robinson & 1966  (11)  &49 & 45(1422)  & 96 & Joe  Hauser & 1924  (3)  &47 \\
42 & Albert  Pujols & 2006  (6)  &49 & 45(1422)  & 96 & Jack  Fournier & 1924  (12)  &47 \\
42 & Mark  McGwire & 1987  (2)  &49 & 45(777)  & 94 & Earl  Averill & 1931  (3)  &47 \\
42 & Willie  Mays & 1962  (11)  &49 & 45(1134)  & 96 & Ken  Williams & 1923  (8)  &47 \\
42 & Ted  Kluszewski & 1954  (8)  &49 & 45(3401)  & 98 & George  Sisler & 1920  (6)  &47 \\
\hline
\end{tabular}}}
\caption{ Ranking of Season Home Runs  for the Modern Era (1920 - 2009). 
}
\label{table:seasonHR}
\end{table}

\begin{table}[h]
\centering{ {\footnotesize
\begin{tabular}{@{\vrule height -.3 pt depth4pt  width0pt}lccc||lcccc}
&\multicolumn4c{{\bf Traditional Rank}}&\multicolumn4c{{\bf Detrended Rank}}\\
\noalign{
\vskip-1pt} Rank & Name & Season ($Y\#$) & Season Metric & Rank$^{*}$(Rank)  &  \% Change & Name & Season ($Y\#$) &
Season Metric \\
\hline 
1 & Ichiro  Suzuki & 2004  (4)  &262 & 1(1)  & 0 & Ichiro  Suzuki & 2004  (4)  &259 \\
2 & George  Sisler & 1920  (6)  &257 & 2(13)  & 84 & Wade  Boggs & 1985  (4)  &247 \\
3 & Bill  Terry & 1930  (8)  &254 & 3(2)  & -50 & George  Sisler & 1920  (6)  &245 \\
3 & Lefty  O'Doul & 1929  (6)  &254 & 4(18)  & 77 & Don  Mattingly & 1986  (5)  &244 \\
5 & Al  Simmons & 1925  (2)  &253 & 5(8)  & 37 & Ty  Cobb & 1911  (7)  &243 \\
6 & Chuck  Klein & 1930  (3)  &250 & 5(27)  & 81 & Kirby  Puckett & 1988  (5)  &243 \\
6 & Rogers  Hornsby & 1922  (8)  &250 & 7(32)  & 78 & Matty  Alou & 1969  (10)  &242 \\
8 & Ty  Cobb & 1911  (7)  &248 & 7(10)  & 30 & Ichiro  Suzuki & 2001  (1)  &242 \\
9 & George  Sisler & 1922  (8)  &246 & 9(16)  & 43 & Rod  Carew & 1977  (11)  &240 \\
10 & Ichiro  Suzuki & 2001  (1)  &242 & 10(36)  & 72 & Joe  Torre & 1971  (12)  &239 \\
11 & Heinie  Manush & 1928  (6)  &241 & 11(44)  & 75 & Nap  Lajoie & 1910  (15)  &237 \\
11 & Babe  Herman & 1930  (5)  &241 & 12(36)  & 66 & Pete  Rose & 1973  (11)  &236 \\
13 & Darin  Erstad & 2000  (5)  &240 & 12(56)  & 78 & Ty  Cobb & 1917  (13)  &236 \\
13 & Jesse  Burkett & 1896  (7)  &240 & 14(18)  & 22 & Ichiro  Suzuki & 2007  (7)  &235 \\
13 & Wade  Boggs & 1985  (4)  &240 & 14(13)  & -7 & Darin  Erstad & 2000  (5)  &235 \\
16 & Willie  Keeler & 1897  (6)  &239 & 16(42)  & 61 & Stan  Musial & 1946  (5)  &233 \\
16 & Rod  Carew & 1977  (11)  &239 & 17(90)  & 81 & Cy  Seymour & 1905  (10)  &232 \\
18 & Ichiro  Suzuki & 2007  (7)  &238 & 18(36)  & 50 & Tommy  Davis & 1962  (4)  &231 \\
18 & Don  Mattingly & 1986  (5)  &238 & 19(119)  & 84 & Ty  Cobb & 1909  (5)  &230 \\
18 & Ed  Delahanty & 1899  (12)  &238 & 19(36)  & 47 & Willie  Wilson & 1980  (5)  &230 \\
21 & Paul  Waner & 1927  (2)  &237 & 19(211)  & 90 & Pete  Rose & 1968  (6)  &230 \\
21 & Joe  Medwick & 1937  (6)  &237 & 19(211)  & 90 & Felipe  Alou & 1968  (11)  &230 \\
21 & Harry  Heilmann & 1921  (7)  &237 & 23(119)  & 80 & Mike  Donlin & 1905  (7)  &229 \\
21 & Hugh  Duffy & 1894  (7)  &237 & 23(68)  & 66 & Kirby  Puckett & 1986  (3)  &229 \\
25 & Jack  Tobin & 1921  (7)  &236 & 25(29)  & 13 & Joe  Jackson & 1911  (4)  &228 \\
26 & Rogers  Hornsby & 1921  (7)  &235 & 25(3)  & -733 & Lefty  O'Doul & 1929  (6)  &228 \\
27 & Lloyd  Waner & 1929  (3)  &234 & 25(150)  & 83 & Nap  Lajoie & 1906  (11)  &228 \\
27 & Kirby  Puckett & 1988  (5)  &234 & 25(101)  & 75 & Pete  Rose & 1969  (7)  &228 \\
29 & Joe  Jackson & 1911  (4)  &233 & 25(101)  & 75 & Felipe  Alou & 1966  (9)  &228 \\
30 & Nap  Lajoie & 1901  (6)  &232 & 25(90)  & 72 & Ralph  Garr & 1971  (4)  &228 \\
30 & Earl  Averill & 1936  (8)  &232 & 25(36)  & 30 & Stan  Musial & 1948  (7)  &228 \\
32 & Freddie  Lindstrom & 1930  (7)  &231 & 32(85)  & 62 & Stan  Musial & 1943  (3)  &227 \\
32 & Freddie  Lindstrom & 1928  (5)  &231 & 32(5)  & -540 & Al  Simmons & 1925  (2)  &227 \\
32 & Earle  Combs & 1927  (4)  &231 & 32(6)  & -433 & Rogers  Hornsby & 1922  (8)  &227 \\
32 & Matty  Alou & 1969  (10)  &231 & 35(30)  & -16 & Nap  Lajoie & 1901  (6)  &226 \\
36 & Willie  Wilson & 1980  (5)  &230 & 35(56)  & 37 & Ichiro  Suzuki & 2009  (9)  &226 \\
36 & Joe  Torre & 1971  (12)  &230 & 35(177)  & 80 & Ty  Cobb & 1907  (3)  &226 \\
36 & Pete  Rose & 1973  (11)  &230 & 38(68)  & 44 & Hank  Aaron & 1959  (6)  &225 \\
36 & Stan  Musial & 1948  (7)  &230 & 38(63)  & 39 & Tommy  Holmes & 1945  (4)  &225 \\
36 & Tommy  Davis & 1962  (4)  &230 & 38(3)  & -1166 & Bill  Terry & 1930  (8)  &225 \\
41 & Rogers  Hornsby & 1929  (15)  &229 & 41(101)  & 59 & Rod  Carew & 1974  (8)  &224 \\
42 & Stan  Musial & 1946  (5)  &228 & 41(113)  & 63 & Tony  Oliva & 1964  (3)  &224 \\
42 & Kiki  Cuyler & 1930  (10)  &228 & 41(11)  & -272 & Heinie  Manush & 1928  (6)  &224 \\
44 & Sam  Rice & 1925  (11)  &227 & 41(21)  & -95 & Joe  Medwick & 1937  (6)  &224 \\
44 & Nap  Lajoie & 1910  (15)  &227 & 45(18)  & -150 & Ed  Delahanty & 1899  (12)  &223 \\
44 & Lance  Johnson & 1996  (10)  &227 & 45(137)  & 67 & Kirby  Puckett & 1989  (6)  &223 \\
44 & Rogers  Hornsby & 1924  (10)  &227 & 45(247)  & 81 & Nap  Lajoie & 1904  (9)  &223 \\
44 & Billy  Herman & 1935  (5)  &227 & 45(119)  & 62 & Paul  Molitor & 1991  (14)  &223 \\
44 & Charlie  Gehringer & 1936  (13)  &227 & 45(9)  & -400 & George  Sisler & 1922  (8)  &223 \\
44 & Jim  Bottomley & 1925  (4)  &227 & 45(229)  & 80 & Roberto  Clemente & 1967  (13)  &223 \\
\hline
\end{tabular}}}
\caption{ Ranking of Season Hits (1890 - 2009).  }
\label{table:seasonH}
\end{table}

\begin{table}[h]
\centering{ {\footnotesize
\begin{tabular}{@{\vrule height .5 pt depth4pt  width0pt}lccc||lcccc}
&\multicolumn4c{{\bf Traditional Rank}}&\multicolumn4c{{\bf Detrended Rank}}\\
\noalign{
\vskip-1pt} Rank & Name & Season ($Y\#$) & Season Metric & Rank$^{*}$(Rank)  &  \% Change & Name & Season ($Y\#$) &
Season Metric \\
\hline 
1 & Hack  Wilson & 1930  (8)  &191 & 1(2)  & 50 & Lou  Gehrig & 1931  (9)  &168 \\
2 & Lou  Gehrig & 1931  (9)  &184 & 2(3)  & 33 & Hank  Greenberg & 1937  (6)  &164 \\
3 & Hank  Greenberg & 1937  (6)  &183 & 3(4)  & 25 & Lou  Gehrig & 1927  (5)  &162 \\
4 & Lou  Gehrig & 1927  (5)  &175 & 4(7)  & 42 & Babe  Ruth & 1921  (8)  &161 \\
4 & Jimmie  Foxx & 1938  (14)  &175 & 5(485)  & 98 & Ty  Cobb & 1907  (3)  &160 \\
6 & Lou  Gehrig & 1930  (8)  &174 & 6(17)  & 64 & Jimmie  Foxx & 1933  (9)  &159 \\
7 & Babe  Ruth & 1921  (8)  &171 & 7(8)  & 12 & Hank  Greenberg & 1935  (4)  &155 \\
8 & Chuck  Klein & 1930  (3)  &170 & 7(4)  & -75 & Jimmie  Foxx & 1938  (14)  &155 \\
8 & Hank  Greenberg & 1935  (4)  &170 & 9(10)  & 10 & Jimmie  Foxx & 1932  (8)  &153 \\
10 & Jimmie  Foxx & 1932  (8)  &169 & 10(16)  & 37 & Babe  Ruth & 1927  (14)  &152 \\
11 & Joe  DiMaggio & 1937  (2)  &167 & 10(1)  & -900 & Hack  Wilson & 1930  (8)  &152 \\
12 & Sam  Thompson & 1895  (11)  &165 & 10(930)  & 98 & Honus  Wagner & 1908  (12)  &152 \\
12 & Al  Simmons & 1930  (7)  &165 & 13(989)  & 98 & Ty  Cobb & 1908  (4)  &150 \\
12 & Manny  Ramirez & 1999  (7)  &165 & 13(350)  & 96 & Sherry  Magee & 1910  (7)  &150 \\
12 & Lou  Gehrig & 1934  (12)  &165 & 13(11)  & -18 & Joe  DiMaggio & 1937  (2)  &150 \\
16 & Babe  Ruth & 1927  (14)  &164 & 16(17)  & 5 & Babe  Ruth & 1931  (18)  &149 \\
17 & Babe  Ruth & 1931  (18)  &163 & 17(126)  & 86 & Joe  Torre & 1971  (12)  &148 \\
17 & Jimmie  Foxx & 1933  (9)  &163 & 17(12)  & -41 & Lou  Gehrig & 1934  (12)  &148 \\
19 & Hal  Trosky & 1936  (4)  &162 & 17(414)  & 95 & Cy  Seymour & 1905  (10)  &148 \\
20 & Sammy  Sosa & 2001  (13)  &160 & 17(1131)  & 98 & Mike  Donlin & 1908  (9)  &148 \\
21 & Hack  Wilson & 1929  (7)  &159 & 17(21)  & 19 & Ted  Williams & 1949  (8)  &148 \\
21 & Ted  Williams & 1949  (8)  &159 & 17(21)  & 19 & Vern  Stephens & 1949  (9)  &148 \\
21 & Vern  Stephens & 1949  (9)  &159 & 23(126)  & 81 & Babe  Ruth & 1920  (7)  &147 \\
21 & Lou  Gehrig & 1937  (15)  &159 & 24(126)  & 80 & Ted  Williams & 1942  (4)  &146 \\
25 & Sammy  Sosa & 1998  (10)  &158 & 24(452)  & 94 & Sam  Crawford & 1910  (12)  &146 \\
26 & Al  Simmons & 1929  (6)  &157 & 24(106)  & 77 & Harmon  Killebrew & 1969  (16)  &146 \\
26 & Juan  Gonzalez & 1998  (10)  &157 & 24(46)  & 47 & George  Foster & 1977  (9)  &146 \\
28 & Alex  Rodriguez & 2007  (14)  &156 & 24(112)  & 78 & Jim  Rice & 1978  (5)  &146 \\
28 & Jimmie  Foxx & 1930  (6)  &156 & 24(232)  & 89 & Gavvy  Cravath & 1913  (4)  &146 \\
30 & Ken  Williams & 1922  (7)  &155 & 24(34)  & 29 & Tommy  Davis & 1962  (4)  &146 \\
30 & Joe  DiMaggio & 1948  (10)  &155 & 31(30)  & -3 & Joe  DiMaggio & 1948  (10)  &145 \\
32 & Babe  Ruth & 1929  (16)  &154 & 31(49)  & 36 & Johnny  Bench & 1970  (4)  &145 \\
32 & Joe  Medwick & 1937  (6)  &154 & 33(30)  & -10 & Ken  Williams & 1922  (7)  &144 \\
34 & Babe  Ruth & 1930  (17)  &153 & 33(63)  & 47 & Don  Mattingly & 1985  (4)  &144 \\
34 & Tommy  Davis & 1962  (4)  &153 & 35(301)  & 88 & Johnny  Bench & 1972  (6)  &143 \\
36 & Rogers  Hornsby & 1922  (8)  &152 & 35(1059)  & 96 & Ty  Cobb & 1909  (5)  &143 \\
36 & Lou  Gehrig & 1936  (14)  &152 & 35(20)  & -75 & Sammy  Sosa & 2001  (13)  &143 \\
36 & Albert  Belle & 1998  (10)  &152 & 35(21)  & -66 & Lou  Gehrig & 1937  (15)  &143 \\
39 & Al  Simmons & 1932  (9)  &151 & 39(232)  & 83 & Bill  Nicholson & 1943  (6)  &142 \\
39 & Mel  Ott & 1929  (4)  &151 & 39(25)  & -56 & Sammy  Sosa & 1998  (10)  &142 \\
39 & Lou  Gehrig & 1932  (10)  &151 & 39(36)  & -8 & Rogers  Hornsby & 1922  (8)  &142 \\
42 & Miguel  Tejada & 2004  (8)  &150 & 42(26)  & -61 & Juan  Gonzalez & 1998  (10)  &141 \\
42 & Babe  Ruth & 1926  (13)  &150 & 42(42)  & 0 & Hank  Greenberg & 1940  (9)  &141 \\
42 & Hank  Greenberg & 1940  (9)  &150 & 42(42)  & 0 & Babe  Ruth & 1926  (13)  &141 \\
42 & Andres  Galarraga & 1996  (12)  &150 & 45(19)  & -136 & Hal  Trosky & 1936  (4)  &140 \\
46 & Ryan  Howard & 2006  (3)  &149 & 45(12)  & -275 & Manny  Ramirez & 1999  (7)  &140 \\
46 & Rogers  Hornsby & 1929  (15)  &149 & 45(28)  & -60 & Alex  Rodriguez & 2007  (14)  &140 \\
46 & George  Foster & 1977  (9)  &149 & 45(194)  & 76 & Enos  Slaughter & 1946  (6)  &140 \\
49 & Rafael  Palmeiro & 1999  (14)  &148 & 45(194)  & 76 & Hank  Aaron & 1963  (10)  &140 \\
49 & David  Ortiz & 2005  (9)  &148 & 50(382)  & 86 & Billy  Williams & 1972  (14)  &139 \\
\hline
\end{tabular}}}
\caption{ Ranking of Season Runs Batted In (RBI), (1890 - 2009).  }
\label{table:seasonRBI}
\end{table}

\begin{table}[h]
\centering{ {\footnotesize
\begin{tabular}{@{\vrule height .5 pt depth4pt  width0pt}lccc||lcccc}
&\multicolumn4c{{\bf Traditional Rank}}&\multicolumn4c{{\bf Detrended Rank}}\\
\noalign{
\vskip-1pt} Rank & Name & Season ($Y\#$) & Season Metric & Rank$^{*}$(Rank)  &  \% Change & Name & Season ($Y\#$) &
Season Metric \\
\hline 
1 & Matt  Kilroy & 1886  (1)  &513 & 1(17)  & 94 & Toad  Ramsey & 1887  (3)  &557 \\
2 & Toad  Ramsey & 1886  (2)  &499 & 2(1)  & -100 & Matt  Kilroy & 1886  (1)  &523 \\
3 & Hugh  Daily & 1884  (3)  &483 & 3(2)  & -50 & Toad  Ramsey & 1886  (2)  &509 \\
4 & Dupee  Shaw & 1884  (2)  &451 & 4(16)  & 75 & Tim  Keefe & 1883  (4)  &478 \\
5 & Charley  Radbourn & 1884  (4)  &441 & 5(13)  & 61 & Mark  Baldwin & 1889  (3)  &463 \\
6 & Charlie  Buffinton & 1884  (3)  &417 & 6(204)  & 97 & Cy  Seymour & 1898  (3)  &457 \\
7 & Guy  Hecker & 1884  (3)  &385 & 6(22)  & 72 & Jim  Whitney & 1883  (3)  &457 \\
8 & Nolan  Ryan & 1973  (7)  &383 & 8(25)  & 68 & Amos  Rusie & 1890  (2)  &445 \\
9 & Sandy  Koufax & 1965  (11)  &382 & 9(129)  & 93 & Dazzy  Vance & 1924  (5)  &443 \\
10 & Bill  Sweeney & 1884  (2)  &374 & 10(3)  & -233 & Hugh  Daily & 1884  (3)  &442 \\
11 & Randy  Johnson & 2001  (14)  &372 & 11(403)  & 97 & Amos  Rusie & 1893  (5)  &439 \\
12 & Pud  Galvin & 1884  (7)  &369 & 12(27)  & 55 & Amos  Rusie & 1891  (3)  &434 \\
13 & Mark  Baldwin & 1889  (3)  &368 & 13(43)  & 69 & Bill  Hutchison & 1892  (5)  &431 \\
14 & Nolan  Ryan & 1974  (8)  &367 & 14(18)  & 22 & Rube  Waddell & 1904  (7)  &419 \\
15 & Randy  Johnson & 1999  (12)  &364 & 15(42)  & 64 & Charley  Radbourn & 1883  (3)  &417 \\
16 & Tim  Keefe & 1883  (4)  &361 & 16(4)  & -300 & Dupee  Shaw & 1884  (2)  &412 \\
17 & Toad  Ramsey & 1887  (3)  &355 & 17(578)  & 97 & Amos  Rusie & 1894  (6)  &410 \\
18 & Rube  Waddell & 1904  (7)  &349 & 18(19)  & 5 & Bob  Feller & 1946  (8)  &407 \\
19 & Bob  Feller & 1946  (8)  &348 & 19(5)  & -280 & Charley  Radbourn & 1884  (4)  &403 \\
20 & Randy  Johnson & 2000  (13)  &347 & 20(29)  & 31 & Tim  Keefe & 1888  (9)  &399 \\
21 & Hardie  Henderson & 1884  (2)  &346 & 21(77)  & 72 & Amos  Rusie & 1892  (4)  &395 \\
22 & Jim  Whitney & 1883  (3)  &345 & 22(496)  & 95 & Amos  Rusie & 1895  (7)  &392 \\
22 & Mickey  Welch & 1884  (5)  &345 & 23(6)  & -283 & Charlie  Buffinton & 1884  (3)  &381 \\
24 & Jim  McCormick & 1884  (7)  &343 & 24(70)  & 65 & Sadie  McMahon & 1890  (2)  &379 \\
25 & Nolan  Ryan & 1977  (11)  &341 & 24(59)  & 59 & Rube  Waddell & 1903  (6)  &379 \\
25 & Amos  Rusie & 1890  (2)  &341 & 26(74)  & 64 & Jack  Stivetts & 1890  (2)  &377 \\
27 & Charlie  Sweeney & 1884  (2)  &337 & 26(74)  & 64 & Bill  Hutchison & 1890  (3)  &377 \\
27 & Amos  Rusie & 1891  (3)  &337 & 28(216)  & 87 & John  Clarkson & 1887  (5)  &372 \\
29 & Tim  Keefe & 1888  (9)  &335 & 29(90)  & 67 & Pud  Galvin & 1883  (6)  &370 \\
30 & Tim  Keefe & 1884  (5)  &334 & 30(299)  & 89 & Dazzy  Vance & 1925  (6)  &368 \\
30 & Randy  Johnson & 2002  (15)  &334 & 31(49)  & 36 & John  Clarkson & 1885  (3)  &366 \\
32 & Nolan  Ryan & 1972  (6)  &329 & 32(44)  & 27 & Walter  Johnson & 1910  (4)  &364 \\
32 & Randy  Johnson & 1998  (11)  &329 & 33(83)  & 60 & John  Clarkson & 1889  (7)  &357 \\
34 & Nolan  Ryan & 1976  (10)  &327 & 34(64)  & 46 & Ed  Morris & 1885  (2)  &354 \\
35 & Ed  Morris & 1886  (3)  &326 & 35(7)  & -400 & Guy  Hecker & 1884  (3)  &352 \\
36 & Tony  Mullane & 1884  (4)  &325 & 36(57)  & 36 & Walter  Johnson & 1912  (6)  &343 \\
36 & Sam  McDowell & 1965  (5)  &325 & 37(10)  & -270 & Bill  Sweeney & 1884  (2)  &342 \\
38 & Lady  Baldwin & 1886  (3)  &323 & 38(887)  & 95 & Doc  McJames & 1898  (4)  &340 \\
39 & Curt  Schilling & 1997  (10)  &319 & 38(79)  & 51 & Bobby  Mathews & 1885  (13)  &340 \\
40 & Sandy  Koufax & 1966  (12)  &317 & 38(321)  & 88 & Matt  Kilroy & 1887  (2)  &340 \\
41 & Curt  Schilling & 2002  (15)  &316 & 41(204)  & 79 & Noodles  Hahn & 1901  (3)  &338 \\
42 & Charley  Radbourn & 1883  (3)  &315 & 41(271)  & 84 & Vic  Willis & 1902  (5)  &338 \\
43 & Bill  Hutchison & 1892  (5)  &314 & 43(12)  & -258 & Pud  Galvin & 1884  (7)  &337 \\
44 & J.R.  Richard & 1979  (9)  &313 & 44(132)  & 66 & Bill  Hutchison & 1891  (4)  &336 \\
44 & Pedro  Martinez & 1999  (8)  &313 & 44(109)  & 59 & Ed  Walsh & 1908  (5)  &336 \\
44 & Walter  Johnson & 1910  (4)  &313 & 46(78)  & 41 & Rube  Waddell & 1905  (8)  &335 \\
44 & John  Clarkson & 1886  (4)  &313 & 46(147)  & 68 & Toad  Ramsey & 1890  (6)  &335 \\
48 & Steve  Carlton & 1972  (8)  &310 & 46(118)  & 61 & Christy  Mathewson & 1903  (4)  &335 \\
49 & Larry  McKeon & 1884  (1)  &308 & 46(14)  & -228 & Nolan  Ryan & 1974  (8)  &335 \\
49 & Mickey  Lolich & 1971  (9)  &308 & 50(137)  & 63 & Bob  Feller & 1941  (6)  &334 \\
\hline
\end{tabular}}}
\caption{ Ranking of Season Strikeouts (1883 - 2009).  }
\label{table:seasonK1}
\end{table}

\begin{table}[h]
\centering{ { \footnotesize
\begin{tabular}{@{\vrule height .5 pt depth4pt  width0pt}lccc||lcccc}
&\multicolumn4c{{\bf Traditional Rank}}&\multicolumn4c{{\bf Detrended Rank}}\\
\noalign{
\vskip-1pt} Rank & Name & Season ($Y\#$) & Season Metric & Rank$^{*}$(Rank)  &  \% Change & Name & Season ($Y\#$) &
Season Metric \\
\hline 
1 & Nolan  Ryan & 1973  (7)  &383 & 1(72)  & 98 & Dazzy  Vance & 1924  (5)  &443 \\
2 & Sandy  Koufax & 1965  (11)  &382 & 2(6)  & 66 & Bob  Feller & 1946  (8)  &407 \\
3 & Randy  Johnson & 2001  (14)  &372 & 3(197)  & 98 & Dazzy  Vance & 1925  (6)  &368 \\
4 & Nolan  Ryan & 1974  (8)  &367 & 4(4)  & 0 & Nolan  Ryan & 1974  (8)  &335 \\
5 & Randy  Johnson & 1999  (12)  &364 & 5(79)  & 93 & Bob  Feller & 1941  (6)  &334 \\
6 & Bob  Feller & 1946  (8)  &348 & 6(1)  & -500 & Nolan  Ryan & 1973  (7)  &333 \\
7 & Randy  Johnson & 2000  (13)  &347 & 7(75)  & 90 & Bob  Feller & 1940  (5)  &325 \\
8 & Nolan  Ryan & 1977  (11)  &341 & 8(133)  & 93 & Van  Mungo & 1936  (6)  &323 \\
9 & Randy  Johnson & 2002  (15)  &334 & 9(47)  & 80 & Hal  Newhouser & 1946  (8)  &322 \\
10 & Nolan  Ryan & 1972  (6)  &329 & 10(102)  & 90 & Bob  Feller & 1939  (4)  &321 \\
10 & Randy  Johnson & 1998  (11)  &329 & 11(435)  & 97 & Lefty  Grove & 1926  (2)  &317 \\
12 & Nolan  Ryan & 1976  (10)  &327 & 12(124)  & 90 & Bob  Feller & 1938  (3)  &316 \\
13 & Sam  McDowell & 1965  (5)  &325 & 12(400)  & 97 & Dazzy  Vance & 1923  (4)  &316 \\
14 & Curt  Schilling & 1997  (10)  &319 & 12(367)  & 96 & Dazzy  Vance & 1928  (9)  &316 \\
15 & Sandy  Koufax & 1966  (12)  &317 & 15(12)  & -25 & Nolan  Ryan & 1976  (10)  &310 \\
16 & Curt  Schilling & 2002  (15)  &316 & 16(8)  & -100 & Nolan  Ryan & 1977  (11)  &301 \\
17 & J.R.  Richard & 1979  (9)  &313 & 17(578)  & 97 & Dazzy  Vance & 1927  (8)  &299 \\
17 & Pedro  Martinez & 1999  (8)  &313 & 18(175)  & 89 & Bobo  Newsom & 1938  (8)  &298 \\
19 & Steve  Carlton & 1972  (8)  &310 & 18(382)  & 95 & Dizzy  Dean & 1933  (3)  &298 \\
20 & Mickey  Lolich & 1971  (9)  &308 & 18(17)  & -5 & J.R.  Richard & 1979  (9)  &298 \\
20 & Randy  Johnson & 1993  (6)  &308 & 21(2)  & -950 & Sandy  Koufax & 1965  (11)  &294 \\
22 & Mike  Scott & 1986  (8)  &306 & 21(251)  & 91 & Hal  Newhouser & 1945  (7)  &294 \\
22 & Sandy  Koufax & 1963  (9)  &306 & 23(269)  & 91 & Lefty  Grove & 1930  (6)  &293 \\
24 & Pedro  Martinez & 1997  (6)  &305 & 24(26)  & 7 & J.R.  Richard & 1978  (8)  &289 \\
25 & Sam  McDowell & 1970  (10)  &304 & 24(600)  & 96 & Lefty  Grove & 1928  (4)  &289 \\
26 & J.R.  Richard & 1978  (8)  &303 & 26(767)  & 96 & Lefty  Grove & 1927  (3)  &282 \\
27 & Nolan  Ryan & 1989  (23)  &301 & 27(484)  & 94 & Dizzy  Dean & 1932  (2)  &274 \\
27 & Vida  Blue & 1971  (3)  &301 & 28(499)  & 94 & Red  Ruffing & 1932  (9)  &273 \\
29 & Curt  Schilling & 1998  (11)  &300 & 29(37)  & 21 & Steve  Carlton & 1980  (16)  &272 \\
30 & Randy  Johnson & 1995  (8)  &294 & 30(449)  & 93 & George  Earnshaw & 1930  (3)  &271 \\
31 & Curt  Schilling & 2001  (14)  &293 & 31(10)  & -210 & Nolan  Ryan & 1972  (6)  &270 \\
32 & Roger  Clemens & 1997  (14)  &292 & 31(528)  & 94 & Lefty  Grove & 1932  (8)  &270 \\
33 & Randy  Johnson & 1997  (10)  &291 & 33(791)  & 95 & Pete  Alexander & 1920  (10)  &269 \\
33 & Roger  Clemens & 1988  (5)  &291 & 34(874)  & 96 & Lefty  Grove & 1929  (5)  &268 \\
35 & Randy  Johnson & 2004  (17)  &290 & 35(1174)  & 97 & Walter  Johnson & 1924  (18)  &267 \\
36 & Tom  Seaver & 1971  (5)  &289 & 36(423)  & 91 & Dizzy  Dean & 1936  (6)  &265 \\
37 & Steve  Carlton & 1982  (18)  &286 & 37(499)  & 92 & Dizzy  Dean & 1935  (5)  &264 \\
37 & Steve  Carlton & 1980  (16)  &286 & 38(964)  & 96 & Pat  Malone & 1929  (2)  &261 \\
39 & Pedro  Martinez & 2000  (9)  &284 & 38(37)  & -2 & Steve  Carlton & 1982  (18)  &261 \\
40 & Tom  Seaver & 1970  (4)  &283 & 38(20)  & -90 & Mickey  Lolich & 1971  (9)  &261 \\
40 & Sam  McDowell & 1968  (8)  &283 & 41(346)  & 88 & Johnny  Vander Meer & 1941  (5)  &260 \\
42 & Denny  McLain & 1968  (6)  &280 & 41(1153)  & 96 & George  Uhle & 1926  (8)  &260 \\
43 & Sam  McDowell & 1969  (9)  &279 & 43(537)  & 91 & Hal  Newhouser & 1944  (6)  &259 \\
44 & Bob  Veale & 1965  (4)  &276 & 44(5)  & -780 & Randy  Johnson & 1999  (12)  &258 \\
44 & John  Smoltz & 1996  (9)  &276 & 45(70)  & 35 & Herb  Score & 1956  (2)  &257 \\
44 & Dwight  Gooden & 1984  (1)  &276 & 46(423)  & 89 & Dizzy  Dean & 1934  (4)  &255 \\
47 & Hal  Newhouser & 1946  (8)  &275 & 46(19)  & -142 & Steve  Carlton & 1972  (8)  &255 \\
47 & Steve  Carlton & 1983  (19)  &275 & 46(27)  & -70 & Vida  Blue & 1971  (3)  &255 \\
49 & Mario  Soto & 1982  (6)  &274 & 49(105)  & 53 & Herb  Score & 1955  (1)  &253 \\
49 & Fergie  Jenkins & 1970  (6)  &274 & 49(729)  & 93 & Lefty  Gomez & 1932  (3)  &253 \\
\hline
\end{tabular}}}
\caption{ Ranking of Season Strikeouts for the Modern Era (1920-2009).  }
\label{table:seasonK2}
\end{table}


\end{widetext}


\begin{thebibliography}{99}
\bibitem{justify} Neglecting the differences between pitchers and fielders does not  change the results of this paper,
as the pdfs of longevity and success maintain their functional form, even if one distinguishes between pitchers and
fielders, as performed in \cite{BB1}. Furthermore, the only affect of pitchers on batting prowess is to effectively
reduce the average by a relatively small percent. Also, in this paper we do not distinguish between complete and
incomplete careers, which also contributes only slightly to the distribution of success across all players.
\end{thebibliography}
\end{document}